\documentclass[aps,prd,showkeys,nofootinbib,superscriptaddress]{revtex4-2}

\usepackage{amsmath,slashed,tensor,amssymb,epsfig,amsthm,bm,graphicx,graphics,slashed}

\newcommand{\G}{\textbf}
\newcommand{\be}{\begin{equation}}\newcommand{\ee}{\end{equation}}
\newcommand{\bea}{\begin{eqnarray}}\newcommand{\eea}{\end{eqnarray}}
\newcommand{\brr}{\begin{array}}\newcommand{\err}{\end{array}}
\newcommand{\bit}{\begin{itemize}}\newcommand{\eit}{\end{itemize}}
\newcommand{\ben}{\begin{enumerate}}\newcommand{\een}{\end{enumerate}}

\newcommand{\bbm}{\begin{bmatrix}}\newcommand{\ebm}{\end{bmatrix}}
\newcommand{\ba}{\begin{array}}
\newcommand{\ea}{\end{array}}

\newcommand{\bthe}{\begin{theorem}} \newcommand{\ethe}{\end{theorem}}
\newcommand{\ble}{\begin{Lemma}} \newcommand{\ele}{\end{Lemma}}

\newcommand{\dr}{\mathrm{d}}
\newcommand{\mpsi}{\boldsymbol{\psi}}

\def\ha{\frac{1}{2}}

\def\intx{\int \! \! \mathrm{d}^3 \textbf{x}}
\def\intk{\int \! \! \mathrm{d}^3 \textbf{k}}

\def\ph{\varphi}
\def\lab{\label}\def\lan{\langle}
\def\lf{\left}

\def\non{\nonumber}\def\pa{\partial}\def\ran{\rangle}
\def\rar{\rightarrow}
\def\ri{\right}
\def\al{\alpha}\def\bt{\beta}\def\ga{\gamma}
\def\de{\delta}\def\De{\Delta}

\def\la{\lambda}\def\La{\Lambda}\def\si{\sigma}
\def\om{\omega}\def\Om{\Omega}

\def\mass{{_{1,2}}}
\def\flav{{e,\mu}}\def\1{{_{1}}}\def\2{{_{2}}}
\def\bk{{\bf {k}}}\def\bx{{\bf {x}}}

\newcommand{\ide}{1\hspace{-1mm}{\rm I}}

\def\noHe0{:\;\!\!\;\!\!:H_e(0):\;\!\!\;\!\!:}
\def\noHm0{:\;\!\!\;\!\!:H_\mu(0):\;\!\!\;\!\!:}
\def\nof{:\;\!\!\;\!\!:}
\def\boldsymbol#1{{\bm #1}}

\def\lab{\label}
\def\lan{\langle}
\def\lf{\left}

\def\non{\nonumber}
\def\pa{\partial}\def\ran{\rangle}
\def\rar{\rightarrow}
\def\ri{\right}

\def\al{\alpha}\def\bt{\beta}\def\ga{\gamma}
\def\de{\delta}\def\De{\Delta}

\def\la{\lambda}
\def\La{\Lambda}\def\si{\sigma}
\def\om{\omega}\def\Om{\Omega}

\def\mass{{_{1,2}}}
\def\flav{{e,\mu}}\def\1{{_{1}}}\def\2{{_{2}}}
\def\nof{:\;\!\!\;\!\!:}
\begin{document}
\title{Neutrino mixing and oscillations in quantum field theory: a comprehensive introduction}

\author{L.~Smaldone}
\email{smaldone@ipnp.mff.cuni.cz}

\affiliation{Institute of Particle and Nuclear Physics, Faculty  of  Mathematics  and  Physics, Charles  University, V  Hole\v{s}ovi\v{c}k\'{a}ch  2, 18000  Praha  8,  Czech  Republic.}

\author{G.~Vitiello}
\email{vitiello@sa.infn.it}

\affiliation{Dipartimento di Fisica, Universit\`a di Salerno, Via Giovanni Paolo II, 132 84084 Fisciano, Italy \& INFN Sezione di Napoli, Gruppo collegato di Salerno, Italy.}

\begin{abstract}
We review some of the main results of the quantum field theoretical approach to neutrino mixing
and oscillations. We show that the quantum field theoretical framework, where flavor vacuum
is defined, permits to give a precise definition of flavor states as eigenstates of (non-conserved) lepton charges.  We obtain the exact oscillation formula which in the relativistic limit reproduces the Pontecorvo oscillation formula and illustrate some of the contradictions arising in the  quantum mechanics approximation. We show that the gauge theory structure underlies the neutrino mixing phenomenon and that there exist entanglement  between mixed neutrinos. The flavor vacuum is found to be an entangled generalized coherent state of $SU(2)$. We also discuss flavor energy uncertainty relations, which imposes a lower bound on the precision of neutrino energy measurements  and we show that the flavor vacuum inescapably emerges in certain classes of models with dynamical symmetry breaking.
\end{abstract}

\vspace{-1mm}

\maketitle
\section{Introduction}
 The study of neutrino mixing and oscillations has attracted much attention in the past decades and at present days in the theoretical \cite{Gribov:1968kq,Bilenky:1975tb,Bilenky:1976yj,Bilenky:1977ne,ALFINITO199591,Blasone:1995zc,BHV99,giunti2007fundamentals,Formaggio2016,torrieri,gango,our1,matsas,Blasone:2019rxl,our3,Grimus:2019hlq,extended,volpe1,Naumov:2021vds,Cabo2021} and experimental \cite{Vogel:2015wua,IceCube:2017lak,PhysRevD.98.030001,Nakano:2020lol,OPERA:2021xtu} research activity. Apart from the interest in the specific phenomenon, it is clear that the understanding of the neutrino mixing and oscillations opens the doors to the physics beyond the Standard Model (SM) \cite{cheng1984gauge}. For example, the simple fact that neutrinos are massive particles requires some corrections to the Lagrangian of the SM where they enter as massless particles \cite{PhysRevLett.19.1264,Salam:1968rm,cheng1984gauge}. Moreover, neutrinos are described by quantum fields and their proper treatment requires the formalism of quantum field theory (QFT), and in particular of those features which make QFT drastically different from quantum mechanics (QM). Quantum fields are indeed mathematically characterized by infinitely many degrees of freedom. This is a key feature of QFT allowing it to escape from the strict dictate of the Stone-von-Neumann theorem \cite{Stone1930,vonNeumann1931}, by which, for systems with a finite number of degrees of freedom, all the representations of the canonical commutation (or anticommutation) relations (CCR, or CAR) are unitarily equivalent, i.e. physically equivalent, as in fact it happens in QM. In QFT the Stone-von-Neumann theorem does not apply exactly because there exist infinitely many degrees of freedom and thus infinitely many unitarily non-equivalent representations of the CCR (or CAR) are allowed to exist in QFT. In other words, under proper boundary conditions, different dynamical regimes may exist, i.e. different {\it phases} of the system, each one described by a different (unitary non-equivalent) representation.

The discovery of such a structural aspect of QFT, in the early 50's of the past century, after a first ``disappointment'', was recognized to be the great richness of QFT \cite{friedrichs1953mathematical,barton1963introduction,berezin1966method,umezawa1982thermo,umezawa1993advanced}. It allows, e.g., the phenomenon of the spontaneous symmetry breaking (SSB)
\cite{Miransky:1994vk}, which constitutes the foundations where the very  same SM rests and is of crucial importance in condensed matter physics (e.g. in superconductivity, ferromagnetism, etc.) and in the study of phase transitions \cite{umezawa1982thermo,blasone2011quantum,Miransky:1994vk}.
The new scenario was soon clarified, for example, by the Haag theorem \cite{Haagqft,haag1996local,Bogolyubov:1990kw} showing that in QFT eigenstates of the full Hamiltonian $H$ (including interaction terms)  do not go to the eigenstates of the free Hamiltonian $H_0$ in the limit of the coupling constant $g$ going to zero. We have nonperturbative physics.

A similar situation occurs in neutrino mixing physics. The mixing transformations are discovered to lead to the flavor vacuum representation which is in fact unitarily inequivalent to the massive neutrino representation \cite{ALFINITO199591,Blasone:1995zc} (see also \cite{PhysRevD.59.113003, Hannabuss:2000hy, PhysRevD.64.013011, PhysRevD.65.096015, Hannabuss:2002cv,Lee:2017cqf}).

Our review in this work is thus limited to the study of the neutrino mixing and oscillations in their natural framework of QFT. We also explicitly show a few of the substantial contradictions and fallacies emerging in the naive perturbative approach. Here, we mention, as a first no-go obstacle to the QM mixing treatment, the Bargmann superselection rule stating that coherent superposition of states with different masses is not allowed in non-relativistic QM \cite{10.2307/1969831,PhysRevLett.87.100405}. In the Appendix \ref{ineqcar} it is shown how fields with different masses and Bogoliubov transformations are related.

The unique possibility offered by QFT of defining the flavor vacuum permits to give a precise definition of flavor states as eigenstates of (non-conserved) lepton charges, thus showing its physical significance. This leads in turn to the exact oscillation formula, which differs from the one obtained in the QM approximation \cite{BHV99}. These aspects are discussed in Section \ref{1bfmixing} for the two-flavor neutrino case. The conclusions also apply to the three-flavor neutrino case, whose formalism is shortly summarized in the Appendix \ref{3flavorAppA}. In the Sections \ref{fallacies} and \ref{lepnum} the contradictions intrinsic to the perturbative QM approach are presented.
In Section \ref{gauge} it is shown that time evolution of mixed (flavored) neutrinos exhibits the  structure of a gauge theory and its nonperturbative character is confirmed.
In Section \ref{Entangl},  the entanglement  between the flavor neutrinos is shortly described, also noticing that the flavor vacuum itself is an entangled $SU(2)$ generalized coherent state.
The flavor-energy uncertainty relations, implicit in the mixing phenomenon due to the flavor oscillations, which impose a lower bound on neutrino energy-measurements precision are reviewed in Section \ref{TEUR} and  in Section \ref{Sec1} the dynamical generation of neutrino mixing and flavor vacuum condensate is discussed. The Section \ref{conclusion} is devoted to concluding remarks.
In the Appendix \ref{DiracEq}, by using first quantization methods applied to the Dirac equation, which however do not exhibit  explicitly {\it the foliation} into the QFT unitarily inequivalent representations,  it is obtained the oscillation formula consistent with the one derived in the QFT formalism.

\section{Neutrino mixing in QFT} \label{1bfmixing}

In this section, the explicit construction of the neutrino flavor eigenstates is reviewed \cite{ALFINITO199591, Blasone:1995zc, PhysRevD.59.113003, Hannabuss:2000hy, PhysRevD.64.013011, PhysRevD.65.096015, Hannabuss:2002cv,Lee:2017cqf}. We will derive the exact  oscillation formula (see Eq.\eqref{oscfor} and \eqref{oscfor2} \cite{Blasone:1995zc,BHV99}). The derivation is based on the  characteristic feature of QFT mentioned in the Introduction, namely, the existence of unitarily inequivalent representations of field algebra (in the present case CAR) \cite{barton1963introduction,berezin1966method,umezawa1982thermo,umezawa1993advanced,Miransky:1994vk,blasone2011quantum} (see also Appendix \ref{ineqcar}).

\subsection{Mixing transformation and flavor vacuum} \label{QFT}
In the following, for simplicity, we consider only two neutrino flavors. Our conclusions can be extended to the case of three neutrinos (see Appendix \ref{3flavorAppA}). It is convenient to start by writing down the mixing transformation
\cite{Gribov:1968kq,Bilenky:1975tb,Bilenky:1976yj,Bilenky:1977ne}:
\be  \label{PontecorvoMix1}
\nu_\si(x) \ = \ \sum_{j} \, U_{\si \, j} \nu_j(x) \, ,    ~~\qquad  \sigma = e, \mu;  ~~j = 1, 2.
\ee
$U$ is the mixing matrix
\be
U \ = \ \begin{pmatrix} \cos \theta & \sin \theta \\ -\sin \theta & \cos \theta \end{pmatrix} \, ,
\ee
where $\theta$ is the mixing angle\footnote{In the case of three flavors, one has the \emph{Pontecorvo--Maki--Nakagawa--Sakata} (PMNS) matrix \cite{10.1143/PTP.28.870} (see Appendix \ref{3flavorAppA} ).}

The fields $\nu_j(x)$, $j=1,2$, $x \equiv {\bf x}, t $ denote the Dirac field operators for massive neutrinos with masses $m_j$:
\begin{eqnarray}
\nu _{j}(x) =  \sum_r \,  \int \!\! \frac{\dr^3 k}{(2 \pi)^{\frac{3}{2}}} \, \left[ u_{{\bf k},j}^{r}(t) \, \alpha _{{\bf k},j}^{r} +  \ v_{-{\bf k},j}^{r}(t)  \,\beta _{-{\bf k},j}^{r\dagger
}\right]  e^{i{\bf k}\cdot {\bf x}}  \, .
\label{fieldex}
\end{eqnarray}
with $u^r_{{\bf k},j}(t) \,= \, e^{- i \om_{\G k,j} t}\, u^r_{{\bf k},j}\;$,
$\;v^r_{{\bf k},j}(t) \,= \, e^{ i \om_{\G k,j} t}\, v^r_{{\bf k},j}$,
 $\om_{\G k,j}=\sqrt{|\G k|^2 + m_j^2}$.   $\al^r_{\G k, j}$ and $\beta _{{\bf k},j}^{r}$ are the annihilation operators for the massive neutrino vacuum state $|0 \rangle_{1,2}$:
\be \label{vacm}
\al^r_{\G k, j}|0 \rangle_{1,2} = 0 = \beta _{{\bf k},j}^{r} |0 \rangle_{1,2} \  .
\ee
The anticommutation relations are, as usual,
\be \label{CAR} \{\nu^{\al}_{i}(x), \nu^{\bt\dag }_{j}(y)\}_{t_x=t_y} =
\de^{3}({\bf x}-{\bf y})
\de _{\al\bt} \de_{ij} \ee
\be  \label{CAR2} \{\al ^r_{{\bf k},i}, \al ^{s\dag }_{{\bf q},j}\} = \de
_{\bf k q}\de _{rs}\de _{ij}  \quad ; \quad \{\bt^r_{{\bf k},i},
\bt^{s\dag }_{{\bf q},j}\} =
\de _{\bf k q} \de _{rs}\de _{ij}, \ee
and the orthonormality and completeness relations are:
\bea \label{orth}
u^{r\dag}_{{\bf k},i} u^{s}_{{\bf k},i} =
v^{r\dag}_{{\bf k},i} v^{s}_{{\bf k},i} = \de_{rs}
\quad, \quad u^{r\dag}_{{\bf k},i} v^{s}_{-{\bf k},i} = 0
\quad, \quad \sum_{r}(u^{r\al*}_{{\bf k},i} u^{r\bt}_{{\bf k},i} +
v^{r\al*}_{-{\bf k},i} v^{r\bt}_{-{\bf k},i}) = \de_{\al\bt}\;.
\eea
Note that  for $i \neq j$ and $m_i \neq m_j$, $\sum_{r,s}  v^{r\dag}_{{\bf k},i} u^{s}_{-{\bf k},j} \neq 0$ and similarly for other spinor products.  We then denote the Fock space for $\nu_1$,
$\nu_2$ by
${\cal H}_\mass = \lf\{ \al_\mass^{\dag}\;,\;
\bt_\mass^{\dag}\;,\; |0\ran_\mass \ri\}.$

The Hamiltonian mass terms for the massive and flavored fields are
\begin{eqnarray}
H_{1,2} &=& m_{1} {\bar \nu} _{1}(x) \nu _{1}(x) + m_{2} {\bar \nu} _{2}(x) \nu _{2}(x) , \\
H_{e, \mu} &=& m_{e} {\bar \nu} _{e}(x) \nu _{e}(x) + m_{\mu} {\bar \nu} _{\mu}(x) \nu _{\mu}(x)  + m_{e \mu} ({\bar \nu} _{e}(x) \nu _{\mu}(x)  + {\bar \nu} _{\mu}(x)
\nu _{e}(x)),
\label{H}
\end{eqnarray}
respectively, with $ m_{e} =
m_{1}\cos^{2}\theta + m_{2} \sin^{2}\theta~$,  $m_{\mu} =
m_{1}\sin^{2}\theta + m_{2} \cos^{2}\theta~$,   $m_{e \mu}
= (1/2)\sin2\theta(m_{2}-m_{1}) = (1/2)\tan 2 \theta \ \delta m$, and
 $\delta m\equiv m_{\mu}-m_e$.

Let us now notice that mixing transformation \eqref{PontecorvoMix1} can be formally rewritten as \cite{ALFINITO199591, Blasone:1995zc}
\bea  \non
\nu_{e}^{\al}(x)  &=& G^{-1}_{\theta}(t)
\nu_{1}^{\al}( x)
G_{\theta}(t) \\
\nu_{\mu}^{\al}(x) &=& G^{-1}_{\theta}(t)
\nu_{2}^{\al}(x) \;
 G_{\theta}(t)
\eea
with the generator given by:
\bea G_\theta(t) & = & \exp[\theta\lf(S_{+}(t) - S_{-}(t)\ri)] \, , \\[2mm]
 S_{+}(t) & \equiv &  \int d^{3}{\bf x}\, \nu_{1}^{\dag}(x) \, \nu_{2}(x) \, , \qquad
S_{-}(t) \ \equiv \ \int d^{3}{\bf x} \,\nu_{2}^{\dag}(x) \, 
\nu_{1}(x) \, .
\eea
In fact, from the above equations we get, e.g., for $\nu_e$
$$\frac{d^2}{d\theta^2}\,\nu^{\al}_{e}\,=
\,-\nu^{\al}_{e}$$
with the initial conditions
$$
\lf.\nu^{\al}_{e}\ri|_{\theta=0}=\nu^{\al}_{1} \quad, \quad
\lf.\frac{d}{d\theta}\nu^{\al}_{e}\ri|_{\theta=0}=\nu^{\al}_{2}
$$
and similarly for $\nu_{\mu}$.

\begin{figure}[htbp]
\begin{center}
\includegraphics*[width=9cm]{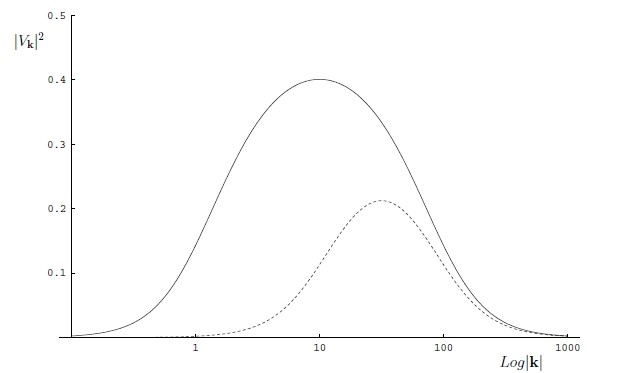}
\end{center}
\caption{$|V_\G k|^2$ for sample values of masses. The solid line corresponds to $m_1=1$ and $m_2=100$, while the dashed line corresponds to $m_1=10$ and $m_2=100$ \cite{blasone2011quantum}.}
	\label{fig:fermioncond}
\end{figure}

The crucial remark is that the vacuum $|0 \ran_\mass$
 is not invariant
under the action of the generator $G_\theta(t)$:
\be \label{timedep}
|0 (t)\ran_\flav \equiv G^{-1}_\theta(t)\; |0 \ran_\mass
=e^{-\theta\lf(S_{+}(t) - S_{-}(t)\ri)}\, |0 \ran_\mass
\ee
The state \eqref{timedep} is known as \emph{flavor vacuum} and it is annihilated by the operators $\al_{\sigma}(t)$ and $\bt_{\sigma}(t)$, defined by:
\be\al_e(t)|0(t)\ran_\flav \ \equiv \ G^{-1}_\theta(t)\al_1
{ G_\theta(t) \; G^{-1}_\theta(t)}|0\ran_\mass \ = \ 0, \ee
and similarly for $\bt_{\sigma}(t)$.
Thus, explicitly,
\bea \lab{operat1}
\al^r_{{\bf k},e}(t)=\cos\theta\,\al^r_{{\bf k},1} +
\sin\theta \lf(
 U_{{\bf k}}^{*}(t)\, \al^r_{{\bf k},2}
 + \epsilon^r
V_{{\bf k}}(t)\, \bt^{r\dag}_{-{\bf k},2}\ri)\quad &&\\
\lab{operat2}
\al^r_{{\bf k},\mu}(t)=\cos\theta\,\al^r_{{\bf
k},2}-\sin\theta \lf(
 U_{{\bf k}}(t)\, \al^r_{{\bf k},1}
 - \epsilon^r
V_{{\bf k}}(t)\, \bt^{r\dag}_{-{\bf k},1}\ri)\quad  &&\\
\lab{operat3}
\!\!\bt^r_{-{\bf k},e}(t)=\cos\theta\,\bt^r_{-{\bf
k},1}+\sin\theta\lf(
U_{{\bf k}}^{*}(t)\, \bt^r_{-{\bf k},2}
 -\epsilon^r
V_{{\bf k}}(t)\, \al^{r\dag}_{{\bf k},2}\ri) \;\; &&\\ \lab{operat4}
\!\!\bt^r_{-{\bf k},\mu}(t)=\cos\theta \, \bt^r_{-{\bf k},2} -
\sin\theta \lf(
 U_{\bf k}(t)\, \bt^r_{-{\bf k},1}  + \epsilon^r
V_{{\bf k}}(t) \, \al^{r\dag}_{{\bf k},1} \ri) \;\; &&
\eea
In Eqs.(\ref{operat1})-(\ref{operat4}), $\epsilon^r \equiv (-1)^r$, and $U_{\bf k}\,$ and $\,V_{\bf k}$ are the \emph{Bogoliubov
coefficients}:
\begin{eqnarray}
U_{{\bf k}}(t)& \equiv & u^{r\dag}_{{\bf k},2}u^r_{{\bf k},1}\;
e^{i(\om_{\G k,2}-\om_{\G k,1})t} \ = \ |U_\G k| \, e^{i(\om_{\G k,2}-\om_{\G k,1})t} \,  \, ,  \\[2mm]
V_{{\bf k}}(t) & \equiv & \epsilon^r\; u^{r\dag}_{{\bf k},1}v^r_{-{\bf k},2}\;
e^{i(\om_{\G k,2}+\om_{\G k,1})t} \ = \ |V_\G k| \, e^{i(\om_{\G k,2}+\om_{\G k,1})t} \, .
\end{eqnarray}
Explicitly
\bea \non
|U_\G k| & \equiv & u^{r\dag}_{{\bf k},2} \, u^{r}_{{\bf k},1} \ = \  v^{r\dag}_{-{\bf k},1} \, v^{r}_{-{\bf k},2} \non \\[2mm]
& = & \left(\frac{\omega_{\G k,1}+m_{1}}{2\omega_{\G k,1}}\right)^{\frac{1}{2}}
\left(\frac{\omega_{\G k,2}+m_{2}}{2\omega_{\G k,2}}\right)^{\frac{1}{2}}
\left(1+\frac{{\bf k}^{2}}{(\omega_{\G k,1}+m_{1})(\omega_{\G k,2}+m_{2})}\right) \, . \label{uk}\\[2mm]
|V_\G k| & = & \epsilon^r\; u^{r\dag}_{{\bf
k},1} \, v^{r}_{-{\bf k},2} \ = \  -\epsilon^r\, u^{r\dag}_{{\bf
k},2} \, v^{r}_{-{\bf k},1} \non \\[2mm]
& = &  \frac{|\G k|}{\sqrt{4 \om_{\G k,1}\om_{\G k,1}}}
\lf(\sqrt{\frac{\om_{\G k,2}+m_2}{\om_{\G k,1}+m_1}}-\sqrt{\frac{\om_{\G k,1}+m_1}{\om_{\G k,2}+m_2}}\ri) \, .
\eea
Notice that $|U_{\bf k}|^2 + |V_{\bf k}|^2 =1$. In the relativistic limit $\omega_{\G k,j} \approx |\G k| $, $|U_\G k| \rightarrow 1$ and $|V_{{\bf k}}| \rightarrow 0$. Also, $|V_{{\bf k}}|=0$  when $m_1=m_2$  and/or $\theta=0$, i.e. when no mixing occurs. $|V_{\bf k}|^2$ has the maximum  at $|\G k|=\sqrt{m_1 m_2}$  with $|V_\G k|^2_{max}\ \rar\  1/2$ for $\frac{( m_{2}-m_{1})^2}{m_{1} m_{2}} \rar \infty$, and $|V_{{\bf k}}|^2\simeq \frac{(m_2 -m_1)^2}{4 |\G k|^2}$
 for $ |\G k|\gg\sqrt{m_1 m_2}$. A plot of $|V_\G k|^2$ is reported in Figure \ref{fig:fermioncond}.

 The flavor fields can be thus expanded as:
\begin{eqnarray} \lab{nue}
\nu_{e}(x)
&=& \sum_{{\bf k},r}  \frac{e^{i {\bf k}\cdot{\bf x}}}{\sqrt{V}}
  \lf[ u^r_{{\bf k},1}(t) \,\al^r_{{\bf k},e}(t)\, +\,
v^r_{-{\bf k},1}(t)\,\bt^{r\dag}_{-{\bf k},e}(t)
\ri] \, ,
\\ [2mm]
 \lab{numu}  \nu_{\mu}(x)
&=& \sum_{{\bf k},r}  \frac{e^{i {\bf k}\cdot{\bf x}}}{\sqrt{V}}
\lf[ u^r_{{\bf k},2}(t)\, \al^r_{{\bf k},\mu}(t)\, + \,
v^r_{-{\bf k},2}(t)\,\bt^{r\dag}_{-{\bf k},\mu}(t)
\ri] \, ,
\end{eqnarray}
and the \emph{flavor Hilbert space} is defined as ${\cal H}_{e,\mu} = \lf\{ \al_{e,\mu}^{\dag}\;,\;
\bt_{e,\mu}^{\dag}\;,\; |0\ran_{e,\mu} \ri\}$.

In order to analyze the explicit structure of flavor vacuum \eqref{timedep}, we decompose the mixing generator as \cite{Blasone:1995zc,BLASONE2016104}:
\bea \label{mavidec}
G_{\theta}  \; = \;  B(m_{1},m_{2})  \;  \; R(\theta)  \; \; B^{-1}(m_{1},m_{2}) \, ,
\eea
where $ B(m_1,m_2)\equiv B_1(m_1) \, B_2(m_2) $, with
\bea
R(\theta) & \equiv & \exp \lf\{\theta  \sum_{{\bf k},r} \Big[\lf(\alpha^{r\dagger}_{{\bf k},1} \alpha^{r}_{{\bf k},2}+ \beta^{r\dagger}_{- {\bf k},1} \beta^{r}_{-{\bf k},2} \ri) e^{i\psi_{\G k}}
- h.c. \Big]\ri\} \, , \\[2mm]
B_i(m_i) & \equiv & \exp{\Big\{  }   \sum_{{\bf k},r} \Theta_{{\bf k},i} \;\epsilon^r  \Big[  \alpha^{r}_{{\bf k},i} \beta^{r}_{-{\bf k},i} e^{-i\phi_{{\G k}\!,\!i}} - \beta^{r\dagger}_{-{\bf k},i} \alpha^{r\dagger}_{{\bf k},i}e^{i\phi_{{\G k},i} }\Big]\Big\} \, , \quad i=1,2 \, .
\eea
Here $\Theta_{\G k,i}= 1/2 \, \cot^{-1}(|\G k|/m_i)$,  $ \psi_{\G k} = (\om_{{\G k},1}-\om_{{\G k},2})t$ and $\phi_{{\G k},i}=2 \om_{{\G k},i}t$. $B_i(\Theta_{{\bf k},i})$, $ i=1,2$ generate the Bogoliubov transformations which are related to mass shifts ($m_1 \neq m_2$) (see Eqs.\eqref{bogfer1},\eqref{bogfer2}), and $ R(\theta)$ generates a rotation. Note that
\be
 R^{-1}(\theta)  |0\ran_{1,2} \ = \ |0\ran_{1,2} \, ,
\ee
and the Bogoliubov transformations induce a condensate structure in the vacuum:
\bea \lab{Bogol}
| \widetilde{0}\ran_{1,2} \ \equiv  \ B^{-1}(m_{1},,m_{2}) |0\ran_{1,2} \ = \ \prod_{{\bf k},r,i} \Big[ \cos{\Theta_{{\bf k},i}} +\epsilon^r \sin{\Theta_{{\bf k},i}} \alpha^{r \dagger}_{{\bf k},i} \beta^{r \dagger}_{-{\bf k},i} \Big] |0\ran_{1,2} \, .
\eea
We remark that the vacuum with condensate  structure Eq. (\ref{Bogol}) is the well known superconductivity vacuum and the (mean-field) vacuum of Nambu--Jona Lasinio model \cite{PhysRev.122.345,PhysRev.124.246}. It is an entangled state for the $\alpha^{r}_{{\bf k},i}$   and $\beta^{r}_{-{\bf k},i}$ modes, for any ${\bf k}, r$ and $i =1,2$.

The decomposition Eq. (\ref{mavidec}) thus shows that \emph{a rotation of fields is not  a simple rotation of creation and annihilation operators}. This fact is clearly exploited by Eqs.(\ref{operat1})-(\ref{operat4}).

Then flavor  vacuum $|0(t) \ran_\flav$  is thus \cite{Blasone:1995zc, BLASONE2016104}  an entangled (see Section \ref{vacuument}) $SU(2)$
generalized coherent state \cite{perelomov2012generalized}. Its explicit condensate structure is given by\footnote{Here we choose $t=0$ as reference time. Moreover we abbreviate $|0\ran_{e ,\mu} \equiv |0(t=0)\ran_{e ,\mu}$. We will often use such abbreviation and $\al^{r\dag}_{\G k,\si} \equiv \al^{r\dag}_{\G k,\si}(0)$.} \cite{Blasone:1995zc}
\begin{eqnarray}
&&{}\hspace{-2cm} \non
|0\ran_\flav= \prod_{{\bf k},r} \lf[
(1-\sin^2\theta\,|V_{{\bf
k}}|^2)
 \ri.-\,\epsilon^r\sin\theta\,\cos\theta\, |V_{{\bf k}}|
\, (\al^{r\dag}_{{\bf k},1}\bt^{r\dag}_{-{\bf k},2}+
\al^{r\dag}_{{\bf k},2} \bt^{r\dag}_{-{\bf k},1})\\ \label{vacuumflav}
&&{}\hspace{-1.6cm}\lf.
+\,\epsilon^r\sin^2\theta \,|V_{{\bf k}}| |U_{{\bf k}}| \,(
\al^{r\dag}_{{\bf k},1}\bt^{r\dag}_{-{\bf k},1} -
\al^{r\dag}_{{\bf k},2}\bt^{r\dag}_{-{\bf k},2} )
+\,\sin^2\theta \, |V_{{\bf k}}|^2
\, \al^{r\dag}_{{\bf k},1}\bt^{r\dag}_{-{\bf k},2}
\al^{r\dag}_{{\bf k},2}\bt^{r\dag}_{-{\bf
k},1}\ri]|0\ran_\mass
\end{eqnarray}
and we see that there are four kinds of condensate
particle-antiparticle pairs with zero momentum and
spin. The condensation density for the $\al^r_{{\bf k},j}$ particle is
\be \lab{condens}
\;_\flav\lan0(t)|\al^{r\dag}_{{\bf k},j}\al^r_{{\bf k},j}
|0(t)\ran_\flav = \sin^2\theta \; |V_{{\bf k}}|^2
\ee
vanishing for $m_\1=m_\2$  and/or $\theta=0$
(in both cases no mixing occurs).
The same result holds for $\bt_j$.

At each time $t$, $|0(t)\ran_\flav$ is orthogonal to the vacuum for the massive neutrinos $|0 \ran_\mass$ in the infinite volume limit $V\rar\infty$ \cite{Blasone:1995zc,PhysRevD.59.113003, Hannabuss:2000hy, PhysRevD.64.013011, PhysRevD.65.096015, Hannabuss:2002cv,Lee:2017cqf}:
\be
\lim_{V \rar \infty}\, _\mass\lan0|0(t)\ran_\flav =
\lim_{V \rar \infty}\, e^{\!V \int
\frac{d^{3}{\bf k}}{(2\pi)^{3}}
\,\ln\,\lf(1- \sin^2\theta\,|V_{\bf k}|^2\ri)^2 }= 0 \label{ineqrep}
\ee
i.e. \emph{flavor and massive fields belong to unitarily inequivalent representations of the CAR}. Moreover, in a similar way, for $t \neq t'$,
\be \label{orthtt}  \lim_{V \rar \infty}\, _\flav \lan0(t')|0(t)\ran_\flav = 0
\ee
In other words, flavor representations at different times are unitarily inequivalent.  This fact reminds us of the quantization of unstable particles \cite{DeFilippo:1977bk,PhysRevD.28.2621}, of QFT in the curved space-time  \cite{Martellini:1978sm}, and of quantum dissipative systems \cite{Celeghini:1991yv}. In particular, we will come back to the formal analogy of neutrinos with unstable particles in Section \ref{TEUR}.

In the Appendix \ref{ineqcar} we further  analyze the Bogoliubov transformations in relation to fields with different masses.  We consider in the following the flavor charge structure and obtain the QFT oscillation formulas.


 \subsection{Flavor eigenstates, charges and neutrino oscillations} \label{bfmixing}

Let us now introduce the flavor charges for a weak decay Lagrangian. Consider, for example, the weak decay reaction $W^+\rightarrow e^+ + \nu_e$. The relevant effective Lagrangian is
$\label{Lagrangian} \mathcal{L}=\mathcal{L}_\nu+\mathcal{L}_l + \mathcal{L}_{int}$ with
\bea \lab{neutr}
&&\mbox{\hspace{-2mm}}{\cal L}_{\nu}  =  \overline{\nu}  \lf( i \ga_\mu \pa^\mu - M_{\nu} \ri)\nu \, ,  \\[2mm]
&&\mbox{\hspace{-2mm}}{\cal L}_{l}  = \overline{l} \lf( i \ga_\mu \pa^\mu - M_{l} \ri) l  \lab{lept} \, , \\[2mm]
 &&\mbox{\hspace{-2mm}}{\cal L}_{int}  =  \frac{g}{2\sqrt{2}}
\lf [ W_{\mu}^{+}\,
\overline{\nu}\,\gamma^{\mu}\,(1-\gamma^{5})\,l +
h.c. \ri] \, ,
\label{Linteract}
\eea
where $\nu$ and $l$ are flavor doublets for neutrinos and charged leptons, while $M_{\nu}$ and $M_{l}$ are the respective mass matrices. In the two-flavor case  $\nu = \lf(\nu_e, \nu_\mu \ri)^{{T}}  , \, l = \lf(e, \mu \ri)^T$, and
\bea \label{neutmass}
&&\mbox{\hspace{-5mm}} M_{\nu}\,=\,  \lf(\ba{cc}m_e & m_{e \mu}
\\ m_{e \mu} & m_\mu\ea\ri) \, , \qquad
M_l\,=\,  \lf(\ba{cc}\tilde{m}_e &0 \\
0 & \tilde{m}_\mu\ea\ri) .
\eea
Of course, the components $m_{e \mu}$ in $M_{\nu}$ imply the presence of the bilinear neutrino mixing terms in ${\cal L}_{\nu}$.
Note that $\mathcal{L}_{\nu}$ can be diagonalized by the mixing transformation Eq. (\ref{PontecorvoMix1}) \cite{Gribov:1968kq,Bilenky:1975tb,Bilenky:1976yj,Bilenky:1977ne,Bilenky:1987ty}, so that
\bea
{\cal L} & =  & \sum_j \, \overline{\nu}_j  \lf( i \ga_\mu \pa^\mu - m_j \ri)\nu_j  \, + \, \sum_\si \, \overline{l} \lf( i \ga_\mu \pa^\mu - \tilde{m}_{\si} \ri) l  \non \\[2mm]
& + &  \frac{g}{2\sqrt{2}} \sum_{\si,j} \,\lf[ W_{\mu}^{+}(x)\, \overline{\nu}_{j} \, U^{*}_{j \si}\,\gamma^{\mu}\,(1-\gamma^{5})\, l_\si +
h.c. \ri] \, .
\eea

The Lagrangian $\mathcal{L}$ is invariant under the global $U(1)$ transformations
$\nu \rightarrow e^{i \alpha} \nu$ and $l \rightarrow e^{i \alpha} l$
leading to the conservation of the total  flavor charge $Q_{l}^{tot}$ corresponding to the lepton-number conservation~\cite{Bilenky:1987ty}. This can be written in terms of the flavor charges for neutrinos and charged leptons~\cite{Blasone:2001qa}
\be
Q_{l}^{tot} =  \sum_{\si=e,\mu} Q_\si^{tot}(t) \,,\quad   Q_{\si}^{tot} (t) = Q_{\nu_{\si}}(t) + Q_{\si}(t)\,,
\ee
with
\bea
Q_{e} & = &  \intx \,
e^{\dag}(x)e(x) \,, \qquad Q_{\nu_{e}} (t) =  \intx \,
\nu_{e}^{\dag}(x)\nu_{e}(x)\,,
\nonumber \\ [2mm]
Q_{\mu} & = &   \intx \,
 \mu^{\dag}(x) \mu(x)\,, \qquad Q_{\nu_{\mu}} (t)= \intx \, \nu_{\mu}^{\dag}(x) \nu_{\mu}(x)\,  .
 \label{QflavLept}
\eea
The above charges can be derived via Noether's theorem~\cite{Blasone:2001qa}  from the Lagrangian~\eqref{Linteract}. Explicitly:
\bea
Q_{\nu_{e}}(t)\!\! &=& \!\! \cos^2\theta\;  Q_{\nu_1} +
\sin^2\theta \; Q_{\nu_2} + \sin\theta\cos\theta \int d^3{\bf x} \lf[\nu_1^\dag
(x) \nu_2(x) + \nu_2^\dag(x) \nu_1(x)\ri]\,, \label{carichemix1}
\\[2mm]
Q_{\nu_{\mu}}(t) \!\! &=& \!\! \sin^{2}\theta \; Q_{\nu_1}
+\cos^{2}\theta \; Q_{\nu_2} - \sin\theta \cos\theta \int d^3{\bf x}
\lf[\nu_1^\dag(x) \nu_2(x) + \nu_2^\dag(x) \nu_1(x)\ri]\,.  \label{carichemix2}
\eea
with $Q_j$, $j = 1,2$, two conserved (Noether) charges :
\bea\label{su2noether}
&&Q_{\nu_j}\, = \,\int d^{3}{\bf x} \,  \nu_{j}^{\dag}(x)\;\nu_{j}(x)\,,
\qquad j=1,2,
\eea
and $Q_1+Q_2 = Q_{\nu_e}(t) + Q_{\nu_\mu}(t)$. Note that contributions in $Q_{\nu_\sigma}$ that cannot be written in terms of $Q_j$ are related to the non-trivial structure of the
flavor Hilbert space.

 By observing that $[\mathcal{L}_{int}({\bf x},t),Q_\si^{tot}(t)]=0$, we see that neutrinos are produced and detected with a definite flavor~\cite{PhysRevD.45.2414, giunti2007fundamentals, BLASONE200937}. However, $[(\mathcal{L}_{\nu} + \mathcal{L}_{l}) ({\bf x},t),Q_\si^{tot}(t)] \neq 0$. In other words, flavor is not preserved by neutrino propagation. This fact gives rise to the phenomenon of flavor oscillations. By using the words of Ref. \cite{close2010neutrino}, when neutrinos are produced and detected they `'\emph{carry identity cards}'', i.e. a definite flavor and `'\emph{can surreptitiously change them if given the right opportunity}''.

The previous discussions suggest that flavor states $|\nu^{r}_{\G k,\si}\ran \ $ are made by flavored particle states obtained according to the natural choice \cite{PhysRevD.60.111302}
\be \label{bvflavstate}
|\al^{r}_{\G k,\si}\ran \ = \ \al^{r\dag}_{\G k,\si} |0\ran_{e ,\mu}  \, .
\ee
and similarly for the antineutrinos\footnote{Again we choose $t=0$ as reference time. Moreover we abbreviated $|0\ran_{e ,\mu} \equiv |0(t=0)\ran_{e ,\mu}$, $\al^{r\dag}_{\G k,\si} \equiv \al^{r\dag}_{\G k,\si}(0)$. We will use often use such abbreviations along the manuscript.} ($|\bt^{r}_{\G k,\si} \ran \equiv \bt^{r\dag}_{\G k,\si} |0\ran_{e ,\mu}$). One can prove that such flavor states are eigenstates of the charge operators:
\be
Q_{\nu_\si}(0) |\nu^r_{\G k,\si}\ran \ = \ |\nu^r_{\G k,\si}\ran \, .
\ee
In the same way, one can build a basis of the flavor Hilbert space, by repeated action of flavor creation operators on the flavor vacuum. 

As first application one can derive oscillation formula by taking the expectation value of the flavor charges~\cite{BHV99}
\be
\mathcal{Q}_{\si\rightarrow \rho}(t) \ = \ \lan Q_{\nu_\rho}(t) \ran_\si \, ,
\ee
where $\langle \cdots\rangle_\si \equiv \lan \nu^r_{\G k,\si}| \cdots |\nu^r_{\G k,\si}\ran$, which gives
\bea \label{oscfor}
&& \mbox{\hspace{-4mm}}\mathcal{Q}_{\si\rightarrow \rho}(t)  =   \sin^2 (2 \theta)\Big[|U_\G k|^2\sin^2\lf(\frac{\Om_{\G k}^{_-}}{2}t\ri)+  |V_\G k|^2\sin^2\lf(\frac{\Om_{\G k}^{_+}}{2}t\ri)\Big]  , \quad \si \neq \rho \, ,  \\[1mm]  \label{oscfor2}
&& \mbox{\hspace{-4mm}}\mathcal{Q}_{\si\rightarrow \si}(t)  =  1 \ - \ \mathcal{Q}_{\si\rightarrow \rho}(t) \, , \quad \si \neq \rho \, ,
\eea
where $\Om_{\G k}^{_{\pm}}\equiv \om_{\G k,2}\pm\om_{\G k,1}$. Notice the presence of the  term proportional to $|V_\G k|^2$ in the oscillation formula Eq. (\ref{oscfor}) (and   (\ref{oscfor2})). We recall that $|V_\G k|^2$ provides a measure of the condensate structure of the flavor vacuum, as shown in the previous subsection \ref{QFT} (see e.g. Eq. (\ref{condens})). As  already mentioned, $|V_\G k|^2 \rar 0$ in the relativistic limit $|{\bf k}| >> m_j$, $j=1,2$, and the oscillation formula reduces to the Pontecorvo one (see Eqs.\eqref{stafor},\eqref{stoscfor} below) in such a limit. As observed in \cite{Blasone:1995zc}, Eq. (\ref{oscfor}) and (\ref{oscfor2}) still give an exact result in the relativistic limit, i.e. although they reproduce the Pontecorvo formulas in that limit, they are not the result of the quantum mechanics finite volume approximation. The dynamical scenario out of which they are derived is not the one of perturbative physics. In this sense, the long debated problem on the building up a Fock space for flavor states \cite{PhysRevD.37.1935,PhysRevD.45.2414,Blasone:1995zc,Ho:2012yja,Lobanov:2015esa,Fantini:2018itu} finds only a {\it formal} agreement on the existence of such a Fock space for relativistic neutrinos. The deep nature of the phenomenon of neutrino mixing and oscillations remains obscure as far as one is trapped in the inappropriate perturbative approach.

In the Sections \ref{fallacies} and \ref{lepnum} we review some of the contradictions emerging in such an inappropriate approach, where the existence of the QFT inequivalent representations of the CAR is ignored.

\section{Contradictions in the quantum mechanics approach to neutrino mixing and oscillations} \label{fallacies}

On the basis of the results obtained in the previous section we now review problems and contradictions arising in the perturbative approximation. We show in this section and in the following one that the definition of flavor states as linear combination of mass-eigenstates leads indeed to some absurd consequences. The main reason is that such states are approximately eigenstates of flavor charges only for high momenta.

We consider first the derivation of the oscillation formula in the QM formalism, then we consider the paradox arising in the study of Green's functions. Then, in the next section, we consider the lepton number conservation in the production vertex.

\subsection{The Pontecorvo oscillation formula}

We consider the states originally introduced as the flavor states by Pontecorvo and collaborators~\cite{Gribov:1968kq,Bilenky:1975tb,Bilenky:1976yj,Bilenky:1977ne,Bilenky:1987ty}:
\be \label{postate}
|\nu^r_{\G k,\si} \ran_{_P} \ = \  \sum_{j} \, U^*_{j \, \si} \, |\nu^r_{\G k,j} \ran \, .
\ee
They are constructed considering the vacuum state  $|0\ran_\mass$ which is annihilated by $\al^{r}_{\G k, j}$ and $\bt^{r}_{\G k, j}$ (the \emph{mass vacuum}) to be the unique vacuum state of the theory.
At fixed time, they are eigenstates of flavor charges only for high momenta:
\be \label{relei}
\lim_{m_i/|\G k|\rightarrow 0} \, Q_{\nu_\si}(0)|\nu^r_{\G k,\si} \ran_{_P} \ = \ |\nu^r_{\G k,\si} \ran_{_P} \, .
\ee
Indeed, this is not true at all energy scales. To see this, we consider the 2-flavor case ($\si=e,\mu$ and $j=1,2$) and evaluate the oscillation formula as the expectation value of the flavor charge on a reference neutrino state \cite{Blasone:2001qa,BHV99,PhysRevD.78.113007}:
\be
\widetilde{\mathcal{P}}_{e\rightarrow \mu}(t) \ \equiv \ {}_{_P}\lan \nu^r_{\G k,e}|Q_{\nu_\mu}(t)|\nu^r_{\G k,e} \ran_{_P} \ = \ \frac{\sin^2 (2 \theta)}{2}\Big[1-|U_\G k| \, \cos \lf(\Om_{\G k}^{_-}t \ri) \Big] \, ,
\ee
where $|U_\G k|$ is given by Eq. (\ref{uk}). For $\om_{\G k,j} >> m_j$, $J=1, \, 2$, it is $|U_\G k| \approx 1$, and we get the Pontecorvo oscillation formula \cite{Gribov:1968kq,Bilenky:1975tb,Bilenky:1976yj,Bilenky:1977ne,Bilenky:1987ty}, $\widetilde{\mathcal{P}}_{e\rightarrow \mu}(t) \rar \mathcal{P}_{e\rightarrow \mu}(t)$:
\be 	\label{stafor}
\mathcal{P}_{e\rightarrow \mu}(t) \ = \ \sin^2 (2 \theta)\,\sin^2 \lf(\frac{\Om_{\G k}^{_-}}{2}t\ri) \, ,
\ee
which can be put in a even more familiar (and experimentally suitable) form
\bea  \label{stoscfor}
\mathcal{P}_{e\rightarrow \mu}(L) \ = \  \sin^2 (2 \theta)\sin^2\lf(\frac{\pi L}{L_{osc}}\ri) \,,  \quad \si \neq \rho,
\eea
where  we used that, in the above limit, $t \approx L$, where $L$ is the traveled distance (e.g. the source-detector distance), and we defined $L_{osc}\equiv 4\pi |\G k| / \delta m_{1,2}^2$, $\delta m_{1,2}^2\equiv m_2^2-m_1^2$. $L_{osc}$ is usually called \emph{oscillation length} \cite{giunti2007fundamentals}.

However, at energies where $|U_\G k| \neq  1$, we have at $t=0$
\be \label{wpf}
\widetilde{\mathcal{P}}_{e\rightarrow \mu}(0) \ = \ \frac{\sin^2 (2 \theta)}{2}\lf(1-|U_\G k|\ri) \, ,
\ee
which is unacceptable because it tells us that flavor is undefined even at $t=0$~\cite{PhysRevD.78.113007,BLASONE200937}. A similar paradox, which does not occur in the case of the exact oscillation formulas Eq. (\ref{oscfor}) and (\ref{oscfor2}), was found in Ref.~\cite{BHV99}, in connection with the study of two-point Green's functions for flavor fields and it will be reviewed in the following.

\subsection{Two-point Green's functions for flavor fields} \label{greensec}
Following Ref.\cite{BHV99} we now  write down the propagator for flavor fields. Some inconsistencies related to the use of the mass vacuum, in the construction of flavor states, emerge also in this context.

Firstly, consider the propagator constructed on the mass vacuum\footnote{In this Section we use the same notation as in Ref.\cite{BHV99}. The propagator is thus defined as the vacuum expectation value of the time-ordered products of fields. Such a definition differs from the one of Ref.~\cite{greiner2013field} just by a factor $i$.} $|0\ran_{1,2}$:
\bea \label{massprop}
S_f(x,y)
& = &
\begin{pmatrix}
 S^{\al \bt}_{e e}(x,y) &  S^{\al \bt}_{e \mu}(x,y) \\ S^{\al \bt}_{\mu e}(x,y) &  S^{\al \bt}_{\mu \mu}(x,y)
\end{pmatrix}
\\ \non
& = & \begin{pmatrix}
{}_{1,2}\lan 0 |T\lf[\nu^\al_e(x) \, \overline{\nu}^\bt_e(y)\ri]|0\ran_{1,2} & \, \,  {}_{1,2}\lan 0 |T\lf[\nu^\al_e(x) \, \overline{\nu}^\bt_\mu(y)\ri]|0\ran_{1,2} \\[2mm] {}_{1,2}\lan 0 |T\lf[\nu^\al_\mu(x) \, \overline{\nu}^\bt_e(y)\ri]|0\ran_{1,2} & \, \,  {}_{1,2}\lan 0 |T\lf[\nu^\al_\mu(x) \, \overline{\nu}^\bt_\mu(y)\ri]|0\ran_{1,2}
\end{pmatrix} \, .
\eea
This can be explicitly written in terms of the propagators of mass fields:
\be
S_f(x,y)
=
\left[
\begin{array}{cc}
 S^{\al \bt}_1(x,y) \cos^2 \theta+S^{\al \bt}_2(x,y) \sin ^2 \theta  &  \, \,  (S^{\al \bt}_2(x,y)-S^{\al \bt}_1(x,y)) \cos \theta \sin \theta  \\
 (S^{\al \bt}_2(x,y)-S^{\al \bt}_1(x,y)) \cos \theta  \sin \theta  &  \, \,   S^{\al \bt}_2(x,y)\cos^2 \theta+S^{\al \bt}_1(x,y) \sin ^2\theta  \\
\end{array}
\right] \, ,
\ee
where
\be
S_j^{\al \bt} \ = \ i \int \! \frac{\dr^4 k}{(2 \pi)^4} \, e^{-i k \cdot (x-y)} \, \frac{\slashed{k}+m_j}{k^2-m^2_j+i \varepsilon} \, , \qquad j=1,2 \, .
\ee
In Ref.\cite{BHV99} it was defined the amplitude of the process where a $\nu_e$ neutrino is created at $t=0$ and its flavor is observed unchanged at time $t>0$, as:
\be
\mathcal{P}^{>}_{ee}(\G k, t) \ = \ i \, u^{r\dag}_{\G k,1} \, e^{i \om_{\G k,1} t} \, S^{>}_{ee}(\G k,t) \, \ga^0 \,  u^{r}_{\G k,1} \, ,
\ee
where $S^{>}_{ee}(\G k,t)$ is the Fourier transform of the Wightman function
\be
S^{>}_{ee}(t,\G x; 0,\G y)= {}_{1,2}\lan 0 |\nu_e(t,\G x) \, \overline{\nu}_e(0,\G y)|0\ran_{1,2} \, .
\ee
An explicit computation gives
\be
\mathcal{P}^{>}_{ee}(\G k, t) \ = \ \cos^2 \theta \ + \ \sin^2 \theta \, |U_\G k|^2 \, e^{-i (\om_{\G k,2}-\om_{\G k,1})t} \, .
\ee
However this result is unacceptable because
\be \label{wic}
\mathcal{P}^{>}_{ee}(\G k, 0^+) \ = \ \cos^2 \theta \ + \ \sin^2 \theta |U_\G k|^2 \ < \ 1\, .
\ee
This inconsistency, which is similar to the one encountered in Eq.\eqref{wpf}, disappears by considering the propagator on the flavor vacuum \cite{BHV99}
\bea \label{flavprop}
\mathcal{G}_f(x,y)
& = &
\begin{pmatrix}
\mathcal{G}^{\al \bt}_{e e}(x,y) & \mathcal{G}^{\al \bt}_{e \mu}(x,y) \\ \mathcal{G}^{\al \bt}_{\mu e}(x,y) & \mathcal{G}^{\al \bt}_{\mu \mu}(x,y)
\end{pmatrix}
\\ \non
& = & \begin{pmatrix}
{}_{e,\mu}\lan 0 |T\lf[\nu^\al_e(x) \, \overline{\nu}^\bt_e(y)\ri]|0 \ran_{e,\mu} &  \, \,   {}_{e,\mu}\lan 0 |T\lf[\nu^\al_e(x) \, \overline{\nu}^\bt_\mu(y)\ri]|0 \ran_{e,\mu} \\[2mm] _{e,\mu}\lan 0 |T\lf[\nu^\al_\mu(x) \, \overline{\nu}^\bt_e(y)\ri]|0\ran_{e,\mu} &   \, \,  {}_{e,\mu}\lan 0|T\lf[\nu^\al_\mu(x) \, \overline{\nu}^\bt_\mu(y)\ri]|0\ran_{e,\mu}
\end{pmatrix} \, ,
\eea
with $y_0=0$. Note that this is indeed consistent from a mathematical point of view. In fact, $|0\ran_{1,2}$ generally does not belong to the domain of $\nu_\si$ (see Eq.\eqref{ineqrep}).

One can check that $\mathcal{G}_f(x,y)$ differs from $S_f(x,y)$ just by boundary terms. For example, we consider the Fourier transform of $\mathcal{G}_{e e}(x,y)$:
\bea \lab{diff}
\mathcal{G}_{e e}(\G k,t) & = & S_{e e}(\G k,t) \ + \ 2 \pi i \sin^2 \theta\lf[|V_\G k|^2(\slashed{k}+m_2)\de(k^2-m^2_2)\ri. \\  & - & |U_\G k||V_\G k| \sum_r\lf(\epsilon^r \, u^r_{\G k, 2}\overline{v}_{-{\bf k},2}^{r}\de(k_0-\om_2)+\epsilon^r v^r_{-\G k, 2}\overline{u}_{{\bf k},2}^{r}\de(k_0+\om_2)\ri)\lf.\ri]  \, .
\eea
By introducing, as before, the Wightman function $\mathcal{G}^{>}_{ee}(\G k,t)$, we define
\be
\mathcal{P}^{>}_{ee}(\G k, t) \ = \ i \, u^{r\dag}_{\G k,1} \, e^{i \om_{\G k,1} t} \, \mathcal{G}^{>}_{ee}(\G k,t) \, \ga^0 \,  u^{r}_{\G k,1} \, .
\ee
This can be evaluated explicitly:
\be
\mathcal{P}^{>}_{ee}(\G k, t) \ = \ \cos^2 \theta \ + \ \sin^2 \theta\lf( |U_\G k|^2 \, e^{-i (\om_{\G k,2}-\om_{\G k,1})t}+|V_\G k|^2 \, e^{i (\om_{\G k,1}+\om_{\G k,2})t}\ri) \, .
\ee
Now $\mathcal{P}^{>}_{ee}(\G k, t)$ satisfies the right initial condition:
\be
\mathcal{P}^{>}_{ee}(\G k, 0^+) \ = \ 1 \, .
\ee
Moreover
\be \label{probcond}
|\mathcal{P}^{>}_{ee}(\G k, t)|^2 \ + \ |\mathcal{P}^{>}_{\mu \mu}(\G k, t)|^2 \ + \ |\mathcal{P}^{>}_{e,\mu}(\G k, t)|^2 \ + \ |\mathcal{P}^{>}_{\mu e}(\G k, t)|^2 \ = \ 1 \, .
\ee
Here $\mathcal{P}^{>}_{\rho\si} \, , \, \rho,\si=e,\mu$ are defined in a similar way as $\mathcal{P}^{>}_{ee}$. We can thus identify the different pieces in the Eq.\eqref{probcond} as flavor (un)-changing probabilities and the QFT oscillation formula can be explicitly derived. The result coincides with Eq.\eqref{oscfor} \cite{BHV99}.

Note that the difference between the propagators on mass and flavor vacuum (cf. Eq.(\ref{diff})) does not appear by using retarded propagators both on flavor and mass vacua \cite{PhysRevD.64.013011}:
\bea
S^{ret}(t,\G x;0,y) & = & \theta(t) \,  {}_{1,2}\lan 0 | \lf\{\nu_\rho(t,\G x),\overline{\nu}_\si(0,\G y)\ri\}|0\ran_{1,2} \, , \\[2mm] \label{retGreen}
\mathcal{G}^{ret}(t,\G x;0,y) & = & \theta(t) \,  {}_{e,\mu}\lan 0 | \lf\{\nu_\rho(t,\G x),\overline{\nu}_\si(0,\G y)\ri\}|0\ran_{e,\mu} \, , \quad \rho,\si=e, \mu \, .
\eea
and
\be \label{eqprop}
S^{ret}(\G k,t) \ = \ \mathcal{G}^{ret}(\G k,t) \, .
\ee
Nevertheless, defining the oscillation probability as
\be \label{yabprob}
\mathcal{Q}_{\nu_\rho \rightarrow \nu_\si}(\G k, t) \ = \ \mathrm{Tr}\lf[\mathcal{G}^{ret}_{\si \rho}(\G k, t)\mathcal{G}^{ret\dag}_{\si \rho}(\G k, t)\ri] \, ,
\ee
one re-obtains Eqs.~\eqref{oscfor} and \eqref{oscfor2} . Once more, this fact shows that the oscillation formulas \eqref{oscfor} and \eqref{oscfor2} are an extremely solid result.

\section{Lepton number conservation in the vertex} \label{lepnum}

In Ref.~\cite{PhysRevD.45.2414} it was pointed out that the amplitude of the neutrino detection process $\nu_\si+X_i \rightarrow e^- + X_f$, where $X_i$ and $X_f$ are the initial and the final particles, respectively, and $e^-$ is the electron, is generally different from zero if $\si \neq e$, if we use the Pontecorvo states. In fact, for low-energy weak processes (where we can use the four-fermion Fermi interaction):
\be \label{wae}
\lan e^{s}_{-}|\bar{e}(x)  \ga^\mu \, (1-\ga^5)  \nu_e(x)|\nu^r_{\si} \ran_{{}_P}  h_\mu(x) \ = \ \sum_j \, U_{e j} 	U^*_{\si j} \lan e^{s}_{-}|\bar{e}(x)  \ga^\mu  (1-\ga^5)  \nu_j(x)|\nu^r_{j} \ran  h_\mu(x)  ,
\ee
where $h_\mu$ are the matrix elements of the $X$ part. This is generally different from $\de_{\si e}$. This seems to be inconsistent because the flavor of the neutrino is \emph{defined} by the flavor of the associated charged lepton in the (lepton-neutrino) doublet (see also Ref. \cite{Bilenky:2001yh}). From this observation, in the attempt of avoiding the inconsistency,
\emph{weak process states} were introduced \cite{PhysRevD.45.2414}.

However, there is an important remark. The $S$ matrix is defined so it connects $in$ and $out$ states \cite{itzykson2012quantum}:
\be
S_{AB} \ \equiv \ \lan  A; out |B; in\ran \, .
\ee
The asymptotic states are defined far before and after the interaction, so that neutrino oscillation occurs leading to the violation of family lepton number conservation. This is actually what it should be expected \cite{Lee:2017cqf}. The previous example is thus not pathological in this sense. However, {\it lepton number has to be conserved in the production and detection vertices} (at tree level), where flavor oscillation can be neglected\footnote{Obviously loop diagrams can produce violation of lepton number in the production and detection vertices, but these contributions are negligible in the following discussion.}. Because this point was debated in literature \cite{Li:2006qt}, it is important to review it in detail.
\subsection{Flavor charge conservation in the vertex}
 Let us consider the (Pontecorvo states) amplitude of the process $W^+\rightarrow e^+ + \nu_e$
\be
\mathcal{A}^P_{W^+ \rightarrow e^+ \, \nu_e} \ = \ {}_{_P}\lan \nu^r_{\G k, e}| \otimes \lan e^s_\G q |\lf[-i \int^{x^0_{out}}_{x^0_{in}} \, \dr^4 x \, \mathcal{H}^e_{int}(x) \ri]|W^+_{\G p, \la} \ran \, .
\ee
The interaction Hamiltonian density is
\be
\mathcal{H}^e_{int}(x) \ = \  -\frac{g}{2\sqrt{2}} \, W^+_\mu(x) \, J^\mu_e(x) + h.c. \, ,
\ee
and
\be
J^\mu_e(x) \ = \ \bar{\nu}_e(x) \, \ga^\mu \, (1-\ga^5)e(x) \, ,
\ee
as it can be deduced from Eq.\eqref{Linteract}. The usual amplitude is obtained by taking the asymptotic limit $x_{out}^0 \rightarrow + \infty$, $x_{in}^0 \rightarrow - \infty$. However, as mentioned, the flavor states are not asymptotic stable states, and we want to investigate the short-time behavior of the amplitude (around the interaction time $x_0=0$). Explicit calculations (cfr. Ref.\cite{Blasone:2006jx}) give:
\bea
\mathcal{A}^P_{W^+ \rightarrow e^+ \, \nu_e} & = & \frac{i g }{2 \sqrt{4 \pi}}\frac{\varepsilon_{\G p, \mu, \la}}{\sqrt{2 E^W_\G p}} \de^3(\G p-\G q -\G k) \non \\[2mm]
& \times & \sum^2_{j=1} \, U^2_{e j}  \int^{x^0_{out}}_{x^0_{in}} \!\! \dr x^0 \, e^{-i \om_{\G k,j} \, x^0_{out}} \, \bar{u}^r_{\G k,j} \, \ga^\mu (1-\ga^5) \, v^s_{\G q,e} \, e^{-i (E^W_\G p-E_q^e-\om_{\G k,j}) x^0}  
\eea
where $E^W_\G p$ and $\varepsilon_{\G p, \mu, \la}$ are the energy and the polarization vector of $W^+$ and $v^s_{\G q,e}$ is the positron wave function. We take $x^0_{in}=-\De t /2$ and $x^0_{out}=\De t /2$, when $\tau_W \ll \De t \ll t_{osc}$, where $\tau_W$ is the $W^+$ lifetime, while $t_{osc}$ is the oscillation time. Under this condition, we can expand the amplitude at the leading order in $\De t$, obtaining:
\bea
\mathcal{A}^P_{W^+ \rightarrow e^+ \, \nu_e} & \approx & \frac{i g }{2 \sqrt{4 \pi}}\frac{\varepsilon_{\G p, \mu, \la}}{\sqrt{2 E^W_\G p}}\, \de^3(\G p-\G q -\G k) \De t \, \sum^2_{j=1} \, U^2_{e j} \ \, \bar{u}^r_{\G k,j} \, \ga^\mu (1-\ga^5) \, v^s_{\G q,e}  \, .
\eea
In the same way, one can evaluate the flavor violating amplitude
\be
\mathcal{A}^P_{W^+ \rightarrow e^+ \, \nu_\mu} \ = \ {}_{_P}\lan \nu^r_{\G k, \mu}| \otimes \lan e^s_\G q |\lf[-i \int^{x^0_{out}}_{x^0_{in}} \, \dr^4 x \, \mathcal{H}^e_{int}(x) \ri]|W^+_{\G p, \la} \ran \, .
\ee
In that case we find
\bea
\mathcal{A}^P_{W^+ \rightarrow e^+ \, \nu_\mu} & = & \frac{i g }{2 \sqrt{4 \pi}}\frac{\varepsilon_{\G p, \mu, \la}}{\sqrt{2 E^W_\G p}}\, \de^3(\G p-\G q -\G k)  \\[2mm] \non
& \times & \sum^2_{j=1} \, U_{\mu j} \, U_{e j} \, \int^{x^0_{out}}_{x^0_{in}} \!\! \dr x^0 \, e^{-i \om_{\G k,j} \, x^0_{out}} \, \bar{u}^r_{\G k,j} \, \ga^\mu (1-\ga^5) \, v^s_{\G q,e} \, e^{-i (E^W_\G p-E_q^e-\om_{\G k,j}) x^0} \, ,
\eea
which in the short time limit becomes
\bea
\mathcal{A}^P_{W^+ \rightarrow e^+ \, \nu_\mu}  \approx  \frac{i g }{2 \sqrt{4 \pi}}\frac{\varepsilon_{\G p, \mu, \la}}{\sqrt{2 E^W_\G p}}\, \de^3(\G p-\G q -\G k) \De t \, \sum^2_{j=1} \, U_{\mu j} \, U_{e j} \ \, \bar{u}^r_{\G k,j} \, \ga^\mu (1-\ga^5) \, v^s_{\G q,e}  \, .
\eea
This is clearly different from zero, which is inconsistent.

The inconsistency disappears when considering the correct QFT flavor states \eqref{bvflavstate}. Let us first evaluate the amplitudes of the decay $W^+ \rightarrow e^+ \, \nu_e$ \cite{Blasone:2006jx}:
\bea \non
\mathcal{A}_{W^+ \rightarrow e^+ \, \nu_e} & = & \frac{i g }{2 \sqrt{2}(2\pi)^\frac{3}{2}} \, \de^3(\G p-\G q -\G k) \, \int^{x^0_{out}}_{x^0_{in}} \!\! \dr x^0 \frac{\varepsilon_{\G p, \mu, \la}}{\sqrt{2 E^W_\G p}}\, \de^3(\G p-\G q -\G k) \\[2mm] \non
& \times &  \lf\{ \cos^2 \theta\, e^{-i \om_{\G k,1} \, x^0_{in}} \, \bar{u}^r_{\G k,1} \, \ga^\mu (1-\ga^5) \, v^s_{\G q,e} \, e^{-i (E^W_\G p-E_q^e-\om_{\G k,1}) x^0} \ri.	\\[2mm]
& + &\sin^2 \theta 	\lf[|U_\G k| \, e^{-i \om_{\G k,2} \, x^0_{in}} \, \bar{u}^r_{\G k,2} \, \ga^\mu (1-\ga^5) \, v^s_{\G q,e} \, e^{-i (E^W_\G p-E_q^e-\om_{\G k,2}) x^0}\ri. \non\\[2mm]
 & + & \lf.\lf. \epsilon^r |V_\G k| \, e^{i \om_{\G k,2} \, x^0_{in}} \, \bar{v}^r_{-\G k,2} \, \ga^\mu (1-\ga^5) \, v^s_{\G q,e} \, e^{-i (E^W_\G p-E_q^e+\om_{\G k,2}) x^0}\ri]\ri\}\, ,
\eea
with $\epsilon^r \equiv (-1)^r$, and then the ``wrong'' one, $W^+ \rightarrow e^+ \, \nu_\mu$:
\bea \non
\mathcal{A}_{W^+ \rightarrow e^+ \, \nu_\mu} & = & \sin \theta \, \cos \theta \, \frac{i g }{2 \sqrt{2}(2\pi)^\frac{3}{2}} \, \de^3(\G p-\G q -\G k) \, \int^{x^0_{out}}_{x^0_{in}} \!\! \dr x^0 \frac{\varepsilon_{\G p, \mu, \la}}{\sqrt{2 E^W_\G p}}\, \de^3(\G p-\G q -\G k) \\[2mm] \non
& \times &  \lf\{ e^{-i \om_{\G k,2} \, x^0_{in}} \, \bar{u}^r_{\G k,2} \, \ga^\mu (1-\ga^5) \, v^s_{\G q,e} \, e^{-i (E^W_\G p-E_q^e-\om_{\G k,2}) x^0} \ri.	 \\[2mm]
& - & 	\lf[|U_\G k| \, e^{-i \om_{\G k,1} \, x^0_{in}} \, \bar{u}^r_{\G k,1} \, \ga^\mu (1-\ga^5) \, v^s_{\G q,e} \, e^{-i (E^W_\G p-E_q^e-\om_{\G k,1}) x^0}\ri. \non\\[2mm]
 & + & \lf.\lf. \epsilon^r |V_\G k| \, e^{i \om_{\G k,1} \, x^0_{in}} \, \bar{v}^r_{-\G k,1} \, \ga^\mu (1-\ga^5) \, v^s_{\G q,e} \, e^{-i (E^W_\G p-E_q^e+\om_{\G k,1}) x^0}\ri]\ri\}\, .
\eea
Looking at the case $\tau_W \ll \De t \ll t_{osc}$, we find~\cite{Blasone:2006jx}:
\bea \non
\mathcal{A}_{W^+ \rightarrow e^+ \, \nu_e} & \approx & \frac{i g }{2 \sqrt{2}(2\pi)^\frac{3}{2}} \, \de^3(\G p-\G q -\G k)  \frac{\varepsilon_{\G p, \mu, \la}}{\sqrt{2 E^W_\G p}}\, \de^3(\G p-\G q -\G k) \, \De t\\[2mm]
& \times &  \lf\{ \cos^2 \theta \, \bar{u}^r_{\G k,2}+\sin^2 \theta 	 \lf[|U_\G k| \bar{u}^r_{\G k,2}+ \epsilon^r |V_\G k| \, \bar{v}^r_{-\G k,2} \ri]\ri\} \, \ga^\mu (1-\ga^5) \, v^s_{\G q,e}\, , 	\label{aee}
\eea
and
\bea 	\label{aemu}
\mathcal{A}_{W^+ \rightarrow e^+ \, \nu_\mu} & \approx & 0\, .
\eea
which is the expected consistent result.

\section{Neutrino mixing and the gauge theory structure} \label{gauge}

 In the flavor basis the kinematic and mass terms in the Lagrangian for the flavored neutrino fields are (cf. Eq. (\ref{neutr}))
\bea \lab{Lagrflav} \mathcal{L}_{\nu}(x)\,=\,  {\bar \nu}(x) \lf( i
\not\!\partial -
  M_\nu \ri) \nu(x) \, .
\eea
For simplicity, when no confusion arises, we omit the space time dependence of the fields, $\nu \equiv \nu(x)$. The field equations derived from $\mathcal{L}_{\nu}(x)$ are
\bea \non  i \partial_0 \nu_e &=& (-i {\boldsymbol{\al}}\cdot{\boldsymbol{\nabla}} + \beta m_e)\nu_e
+ \beta m_{e \mu} \nu_{\mu} \\[2mm]  i \partial_0 \nu_{\mu} &=& (-i
{\boldsymbol{\al}}\cdot{\boldsymbol{\nabla}} + \beta m_{\mu})\nu_{\mu} + \beta m_{e \mu}
\nu_e,  \label{gaugeq}\eea
$\alpha_i$, $i=1,2,3$ and $\beta$ are the Dirac
matrices. We choose the representation
\bea \alpha_i=\lf(\ba{cc}0&\sigma_i\\\sigma_i&0\ea\ri),\qquad
\beta= \lf(\ba{cc}\ide_2 &0\\0&-\ide_2 \ea\ri) \, ,
\eea
where $\si_i$ are the Pauli matrices and $\ide_n$ is the $n \times n$ identity matrix.
In a compact form:
\bea iD_0 \nu= (-i{\boldsymbol{\al}}\cdot{\boldsymbol{\nabla}} + \beta M_d)\nu,\eea
with the diagonal mass matrix $M_d={\rm diag}(m_e,m_{\mu})$.
The mixing term, proportional to $m_{e \mu}$, is taken into account by
the (non-Abelian) covariant derivative \cite{Blasone:2010zn}:
\bea \label{covdevmix}
D_0 \equiv \partial _0 + i\, m_{e \mu}\,\beta\,\sigma_1 . \eea

We thus recognize that flavor  mixing can be described as an interaction of the flavor fields
with an $SU(2)$ constant gauge field:
\bea \label{Conn}
A_{\mu} &\equiv&\frac{1}{2} A_{\mu}^a  \si_a \,=\,  n_{\mu} \delta m \,\frac{\sigma_1}{2}\in
su(2), \qquad n^{\mu}\equiv(1,0,0,0)^T ,\eea
i.e., having only the temporal component in spacetime and only the
first component in $su(2)$ space, $A_{0}^{1} = \delta m$. The
covariant derivative can be written in the form:
\bea D_{\mu}= \partial_{\mu} + i\, g\, \beta\, A_{\mu}, \eea
where we have defined $g\equiv\tan 2\theta$  as the coupling constant for
the mixing interaction. The Lagrangian $\mathcal{L}_{\nu}$ is thus written as
\bea \label{LagrflavCov}\mathcal{L}_{\nu} = {\bar \nu}
(i\gamma^{\mu}D_{\mu} - M_d) \nu. \eea

In the case of maximal mixing
($\theta=\pi/4$), the coupling constant grows to infinity while
$\delta m$ goes to zero.  Since  the gauge
connection is  a constant, with just one non-zero component in
group space, its field strength vanishes identically:
\bea F_{\mu\nu}^a=\epsilon^{abc}A_{\mu}^b A_{\nu}^c=0, ~~~~a,b,c=1,2,3.\eea
Despite $F_{\mu\nu}$ vanishes identically,
the gauge field leads to observable effects (e.g. neutrino oscillations), in analogy with the
Aharonov--Bohm effect \cite{PhysRev.115.485}.

If one goes ahead with the idea of considering the gauge field $A_\mu$, as an `'external field'', it is natural to define the energy momentum tensor as:
\bea \label{newemtensor}\widetilde{T}_{\rho\sigma} = {\bar \nu} i
\gamma_{\rho}D_{\sigma}\nu - \eta _{\rho\sigma}{\bar \nu}
(i\gamma^{\lambda}D_{\lambda} - M_d)\nu.\eea
$\eta_{\rho\sigma}=\textrm{diag}(+1,-1,-1,-1)$ is the Minkowskian metric
tensor. $\widetilde{T}_{\rho\sigma}$ is to be compared with the canonical energy momentum tensor (so that $H=\intx T_{0 0}$):
\bea \label{emtensor}
 T_{\rho\sigma}&=& {\bar \nu} i
\gamma_{\rho}D_{\sigma}\nu - \eta _{\rho\sigma}{\bar \nu}
(i\gamma^{\lambda}D_{\lambda} - M_d)\nu
 + \eta_{\rho\sigma}m_{e \mu} {\bar \nu} \si_1\nu .
\eea
The difference between the two is just the presence of
the interaction terms, proportional to $m_{e \mu}$,  in the $00$ component, i.e. $T_{00}-\widetilde{T}_{00} =
m_{e \mu}({\bar \nu}_e \nu_{\mu} +  {\bar
\nu}_{\mu} \nu_e) $, while we have $T_{0i}=\widetilde{T}_{0i}$, $T_{ij}=\widetilde{T}_{ij}$.
One then may proceed  by defining the $4$-momentum operator
$\widetilde{P}^\mu \equiv \int d^3\mathbf{x} \,\widetilde{T}^{0\mu}$ and the Lorentz group  generators $\widetilde{M}^{\lambda\rho}(x_0)$, which satisfy the usual Poincar\'e algebra at fixed time \cite{Blasone:2010zn}.

What is the physical interpretation of the tilde operators? An intriguing suggestion comes from the possibility of viewing a gauge field as a thermal reservoir \cite{CELEGHINI199298,CELEGHINI1993121,PhysRevA.74.022105,blasone2011quantum}. Then, in Ref. \cite{Blasone:2010zn}, the tilde-Hamiltonian operator $\widetilde{H} \equiv \widetilde{P}^0=\sum_\si \intx \, \nu^\dag_\si \, (-i {\boldsymbol{\al}}\cdot{\boldsymbol{\nabla}} + \beta m_\si) \,  \nu_\si$ was interpreted as a Gibbs free
energy $F \equiv \widetilde{H} = H - T\, S$, where the temperature $T$ is identified with the coupling constant $g=\tan 2\theta$, and entropy $S$ associated to flavor mixing is given by \cite{Blasone:2010zn}
\bea \label{entropy}
 S &=&  \int  d^3\mathbf{x}\,
{\bar \nu} \, A_0 \, \nu\,=\,\frac{1}{2}\, \delta m \int  d^3\mathbf{x}\,
({\bar \nu}_e \nu_{\mu} +  {\bar
\nu}_{\mu} \nu_e) \, .
\eea
with $\delta m\equiv m_{\mu}-m_e$ (cf. Eq. (\ref{H})).  The appearance of an entropy confirms that each one of
the two flavor neutrinos can be viewed  formally as an ``open system'' which
presents some kind of (cyclic) dissipation. This interpretation can be compared with the one presented in Section \ref{TEUR}, where neutrinos are viewed as (cyclic) unstable particles. 

Summing up, the gauge field structure in the mixed neutrino evolution shows that one flavored neutrino evolution is intrinsically dependent on the other flavored neutrino evolution, cf. Eqs. (\ref{gaugeq}). This signals
that an entanglement is present in neutrino mixing, as it will be discussed in the following Section \ref{Entangl}.

\section{Entanglement and neutrino mixing in QFT} \label{Entangl}

The entanglement phenomenon represents a crucial feature of QM and QFT \cite{nielsen2000quantum}. It has been experimentally tested in recent decades and many efforts are devoted to its study  in quantum optics an in general quantum theoretical computing.  It is much interesting that it is also present in neutrino mixing and oscillations \cite{ill1,ill2,ill3,BLASONE2013320,Blasone:2014jea, Bittencourt:2014pda,ill4,ill5}. Here we summarize briefly its main aspects in the QFT neutrino mixing framework. In the subsection \ref{neutrEnt} we consider the entanglement associated to the variances of the observables flavor charges discussed in Section \ref{bfmixing}. In the  subsection \ref{vacuument} we consider the entanglement as a property of the vacuum for the flavor neutrinos, namely the entanglement of the neutrino-antineutrino pairs condensed in it, therefore entanglement as a specific QFT feature of the neutrino mixing.

\subsection{Neutrino entanglement} \label{neutrEnt}

It has been shown that one can regard QM flavor states as entangled states by introducing the
correspondence with two-qubit states \cite{Blasone:2010ta,blasone2011quantum}:
\begin{eqnarray}\label{massqubits}
|\nu_{1}\rangle \equiv |1\rangle_{1} |0\rangle_{2}  \equiv
|10\rangle, \quad
|\nu_{2}\rangle \equiv |0\rangle_{1} |1\rangle_{2}
 \equiv |01\rangle,
\end{eqnarray}
where $|\rangle_{j}$ denotes states in the Hilbert space for neutrinos
with mass $m_j$.
Thus, the occupation number allows to interpret the flavor
states as arising from the entanglement of the $\nu_{j}$, $j=1,2$ neutrinos. This is known as \emph{static} entanglement \cite{Blasone:2010ta,blasone2011quantum}.

In a similar way. a correspondence between two flavor QM states and two-qubit states can be introduced~\cite{ill2,BLASONE2013320,ill5}
\begin{eqnarray}\label{flavorqubits}
 |\nu_{e}\rangle \equiv |1\rangle_{e} |0\rangle_{\mu} ,
 \qquad
|\nu_{\mu}\rangle \equiv |0\rangle_{e} |1\rangle_{\mu} ,
\end{eqnarray}
where  $|0\rangle_{\si}$ and $|1\rangle_{\si}$ denote the absence and the presence of a neutrino in mode $\si$,  respectively. Because it relies on the time evolution of flavor states this phenomenon is known as \emph{dynamic entanglement}.

In QFT, both static and dynamic entanglement are present and can be efficiently quantified by considering the variances of  operator charges \cite{Blasone:2014jea} (see also the general discussion of Ref. \cite{PhysRevA.75.032315}.). For example. considering  the charges $Q_{\nu_j}$, cf. Eq. \eqref{su2noether}, we obtain a measure of the static entanglement present in the states $|\nu^r_{\G k,\si}\ran$:
\bea \label{DEQn1}
\si^2_{Q_{j}} \ \equiv \ \lf(\De Q_{\nu_j}\ri)^2  \ = \ \lan Q_{\nu_j}^2\ran_\rho \,-\,
\lan Q_{\nu_j}\rangle_\rho^2
\ = \  \frac{1}{4} \sin^2(2\theta)\, .
\eea
This result coincides with the QM one \cite{Blasone:2014jea}.
On the other hand, the \emph{dynamic} entanglement can be measured by the variances of the flavor
charges:
\bea \label{varq}
\sigma^2_Q \ \equiv \ \lf(\De Q_{\nu_\rho}\ri)^2  \ = \ \lan Q^2_{\nu_\rho}(t)\ran_\rho \ - \ \lan Q_{\nu_\rho}(t)\ran_\rho^2 \  = \ \ \mathcal{Q}_{\rho \rightarrow \rho}(t)\lf(1-\mathcal{Q}_{\rho\rightarrow \rho}(t)\ri) \, .
\eea
The differences with the relativistic limit result must be traced to the presence of the vacuum condensate affecting the oscillation formulas Eqs. (\ref{oscfor}) and (\ref{oscfor2}).

It is interesting to investigate the origin of the static and the dynamical entanglement in relation with unitarily inequivalent representations into play: in the static entanglement, the flavor Hilbert space at time $t$ to which the entangled state $|\nu_\si(t)\ran$ belongs, is unitarily inequivalent to the Hilbert space for the qubit states $|\nu_i\ran$, see Eq.~(\ref{ineqrep}); on the other hand, in the case of  dynamical entanglement, where the qubits are taken to be  the flavor states at time $t=0$, the inequivalence is among the flavor Hilbert space at different times (cf.  Eq.(\ref{orthtt})).

\subsection{Entangled vacuum state} \label{vacuument}

Particle-antiparticle pairs (${\al^{r}_{{\bf k},i}}, {\bt^{r}_{-{\bf k},j}} $) with zero momentum and
spin appear to be condensate in the flavor vacuum, as shown in Eq. (\ref{vacuumflav}). They are entangled pairs of condensed quanta. The entanglement correlation is due to the coherent condensate structure of the vacuum $|0 \ran_\flav$ generated by the Bogoliubov transformations. A measure of the particle entanglement in $|0 \ran_\flav$
is provided by the linear correlation coefficient $J(N_{a} ,N_{b})$  given by (see e.g. \cite{gerry2005introductory,app9153203})
\be    \lab{corr}
J(N_{a} ,N_{b})  = \frac{cov (N_{a} ,N_{b} )}{(\langle(\Delta N_{a})^{2}\rangle)^{1/2} \, (\langle(\Delta N_{b})^{2}\rangle)^{1/2}} ,
\ee
where $a$ and $b$ generically denote $\al$ and/or $\bt$ quanta,  the symbol $\langle \ldots   \rangle$ denotes expectation value in $|0 \ran_\flav$, $N_{a}$, $N_{b}$  number operators, the variance  $\langle(\Delta{N})^{2}\rangle
 \equiv \langle \lf({N} - \langle {N}\rangle \ri)^{2}  \rangle
= \langle {N}^{2}\rangle - {\langle {N}\rangle}^{2}$,  the covariance  $cov \lf(N_{a} ,N_{b} \ri) \equiv \langle N_{a} N_{b}   \rangle - \langle N_{a}  \rangle \, \langle  N_{b}   \rangle$. 

For non-correlated modes $\langle N_{a} N_{b}   \rangle  = \langle N_{a}  \rangle \, \langle  N_{b}   \rangle$, and $cov \lf(N_{a} ,N_{b} \ri)$ is zero. Moreover, in the case $a =b$, $cov (N_{a} ,N_{a}) = \lan(\Delta N_{a})^{2}\ran$, so that $J=1$. 

Notice that  $\lan(\Delta N_{a})^2\ran$ and/or $\lan(\Delta N_{b})^2\ran$ cannot be zero in Eq. (\ref{corr}), i.e. such values are excluded from the existence domain of 
$J(N_{a} ,N_{b})$. From inspection of the vacuum structure (Eq. (\ref{vacuumflav})) and of the operator transformations (Eq. (\ref{operat1}) - (\ref{operat4})), we find, for $i = 1,2$,  $\lan (\Delta N_{\al^{r}_{{\bf k},i}})^{2}\ran = \sin^{2} \theta \, |V_\G k|^{2} (1-\sin^{2} \theta \, |V_\G k|^{2})$ (and similarly for  $\lan (\Delta N_{\bt^{r}_{{\bf k},i}})^{2}\ran$), which thus has to be not zero. Therefore, we must exclude the values $\theta =0$ and  $|V_\G k|^2 = 0$ ($|U_\G k|^2 =1$), solutions of $ \sin^{2} \theta \, |V_\G k|^{2} = 0$,  and $\theta =\pi /2$ and $|V_\G k|^2 =1$ ($|U_\G k|^2 =0$) where $\sin^{2} \theta = 1/|V_\G k|^2$ can be only satisfied (otherwise it is never satisfied since $1/|V_\G k|^2$ is always larger than 1). 

Summing up,  the definition of $J(N_{\al^{r}_{{\bf k},i}} ,N_{\bt^{r}_{{\bf k},i}})$ is meaningful only within its existence domain, to which the values $\theta =0$, $\theta =\pi /2$, $|V_\G k|^2 = 0, 1$ (i.e. $|U_\G k|^2 =1, 0$), for any $\bf k$, do not belong.

Then, within such existence domain, we find, 
for any $\bf k$  and  $i,j = 1,2$,
\be
J \lf(N_{\al^{r}_{{\bf k},i}} ,N_{\bt^{r}_{-{\bf k},j}}\ri) \ = \ \frac{1}{1+\tan^2 \theta \, |U_\G k|^2}  \, , \qquad i \neq j ,
\ee
and for the pairs (${\al^{r}_{{\bf k},i}}, {\bt^{r}_{-{\bf k},i}}$), $i = 1,2$,
\be
J \lf(N_{\al^{r}_{{\bf k},i}} ,N_{\bt^{r}_{-{\bf k},i}}\ri) \ = \ \frac{|U_\G k|^2 \, \tan^2 \theta}{1+\tan^2 \theta \, |U_\G k|^2} .
\ee

Finally,  $\langle N_{\al^{r}_{{\bf k},i}} N_{\al^{r}_{{\bf k},j} } \rangle = sin^{4} \theta \, |V|^{4} = \langle N_{\al^{r}_{{\bf k},i}}  \rangle \, \langle  N_{\al^{r}_{{\bf k},j}}   \rangle $,  and similarly for $\{ \bt^{r}_{{\bf k},i}, \bt^{r}_{{\bf k},j} \}$ pairs, for  $i \neq j; i, j = 1,2$.  Therefore, $J = 0$ in such cases. 

As we see, the QFT formalism describes and allows to quantify the vacuum entanglement in the neutrino mixing and oscillation phenomenon. 

\section{Flavor--Energy uncertainty relations} \label{TEUR}
In Section \ref{bfmixing} we remarked that
flavor charges do not commute with the free part of the Lagrangian
and then with the corresponding Hamiltonian $H$. This leads to a \emph{flavor-energy uncertainty relation} \cite{Blasone2019},  that can be formalized via the Mandelstam--Tamm \emph{time-energy uncertainty relation} (TEUR) \cite{ManTam}, where flavor charges play the role of clock observables for the oscillating neutrino systems. Note that, because only flavor can be detected in weak processes, such uncertainty relations put a fundamental bound on neutrino energy/mass precision. As remarked in  Refs. \cite{Bilenky2008,Akhmedov:2008zz,Bilenky:2009zz}, Mossbauer neutrinos could furnish an example of neutrinos produced with a definite energy and then this could spoil the oscillation phenomenon. Here we do not discuss such a problem.
\subsection{Time-energy uncertainty relations for neutrino oscillations: Pontecorvo flavor states}
Mandelstam--Tamm version of TEUR is formulated as~\cite{ManTam}
\be \label{teunc}
\Delta E \, \Delta t \, \geq \frac{1}{2} \, .
\ee
We put
\be
\Delta E \equiv \si_H \, \qquad \Delta t \equiv \si_O/\lf|\frac{\dr \lan O(t) \ran}{\dr t}\ri| \, .
\label{teunc1}
\ee
Here $O(t)$ represents the ``clock observable'' whose dynamics quantifies temporal changes in a system
and $\Delta t$ is the characteristic time interval over which the mean value of $O$ changes by a standard deviation.

TEUR for neutrino oscillations in flat spacetime have been extensively studied in Refs. \cite{Bilenky:2005hv,Bilenky2008,Akhmedov:2008zz,Bilenky:2009zz} and later extended to stationary curved spacetimes \cite{Blasone2020}.
Here we follow in our presentation a slightly different strategy.
We start by considering the number operator for flavor (Pontecorvo) neutrinos:
\be
N_{P,\si}(t) \ = \ \sum_{\G k,r} \al^{r\dag}_{P,\G k, \si}(t) \al^r_{P,\G k, \si}(t) \,,
\ee
where the Pontecorvo ladder operators $\al^r_{P,\G k, \si}$  are defined as
\be \label{tiop}
\al^r_{P,\G k, \si} \ = \ \sum^2_{j=1} \, U_{\si \, j} \, \al^r_{\G k, j} \, ,
\ee
with $\al^r_{\G k, j}$ the annihilation operator of fields with definite mass (cf. Eq. (\ref{PontecorvoMix1}) and Eq. (\ref{fieldex})).

The oscillation formula \eqref{stafor}~\cite{Gribov:1968kq,Bilenky:1975tb,Bilenky:1976yj,Bilenky:1977ne,Bilenky:1987ty}  can be obtained by taking the expectation value of the number operator
over Pontecorvo flavor states \eqref{postate}. In particular
\begin{eqnarray}\label{pontosc}
\mathcal{P}_{\si\rightarrow \si}(t)\ = \ \lan N_{P,\si}(t) \ran_\sigma  \ = \  1 - \sin^2(2\theta)\sin^2 \lf(\frac{\Om_{\G k}^{_-}}{2}t\ri) \, ,
\end{eqnarray}
where, in this subsection, $\langle \cdots\rangle_\sigma = {}_{P}\lan \nu^r_{\G k,\si}| \cdots |\nu^r_{\G k,\si}\ran_P$. By setting $O(t)={N_{P,\si}(t)}$
in (\ref{teunc1}) and taking into account that
\bea \non
\si^2_N \ = \  \lan N^2_{P,\si}(t) \ran_\sigma \ - \ \lan N_{P,\si}(t) \ran_\si^2
\ = \ \ \mathcal{P}_{\si\rightarrow \si}(t)\lf(1-\mathcal{P}_{\si\rightarrow \si}(t)\ri) \, ,
\eea
one gets
\be \label{neutunqm}
\lf|\frac{\dr \mathcal{P}_{\si\rightarrow \si}(t)}{\dr t}\ri| \,\leq  \,2\Delta E  \,\sqrt{\mathcal{P}_{\si\rightarrow \si}(t)\lf(1-\mathcal{P}_{\si\rightarrow \si}(t)\ri)} \, .
\ee
 Note that, as discussed in Section \ref{Entangl}, $\si^2_N$ quantifies the dynamic flavor entanglement of the neutrino state \eqref{postate}.

If we consider $\mathcal{P}_{\si\rightarrow \si}(t)$ in the interval $0 \leq t \leq t_{1min}$,
where $t_{1min}$ is the time when $\mathcal{P}_{\si\rightarrow \si}(t)$ reaches the first minimum,
this is a monotonically decreasing function \cite{Bilenky:2005hv,Bilenky2008,Akhmedov:2008zz,Bilenky:2009zz}. In other words, if we try to reveal neutrinos in processes with time scales much smaller than oscillation time, they can be thought as unstable particles \cite{Blasone2019,Blasone:2020qbo}.

The inequality \eqref{neutunqm} can be further simplified by noticing that the maximum value of the r.h.s. is $\De E$, then the weaker inequality
\be
\Delta E  \ \geq  \ \lf|\frac{\dr\mathcal{P}_{\si\rightarrow \si}(t)}{\dr t} \ri| \, ,
\ee
holds. By means of triangular inequality and integrating both
sides from $0$ to $T$, we get
\be
\mbox{\hspace{-2mm}}\Delta E \, T  \ \geq  \ \int^T_{0} \!\! \dr t \, \lf|\frac{\dr\mathcal{P}_{\si\rightarrow \si}(t)}{\dr t} \ri| \, \geq  \,\lf|\int^T_{0} \!\! \dr t \, \frac{\dr\mathcal{P}_{\si\rightarrow \si}(t)}{\dr t} \ri| \, .
\ee
Therefore, one finds
\be \label{etrel}
\Delta E\,  T \geq \mathcal{P}_{\si\rightarrow \rho}(T)  \, ,  \quad \si \neq \rho  \, ,
\ee
with $
\mathcal{P}_{\si\rightarrow \rho}(t) \ = \  1-\mathcal{P}_{\si\rightarrow \si}(t) $.
For $T=T_h$, so that $\mathcal{P}_{\si\rightarrow \rho}(T_h)=\ha$, we finally get
\be
\Delta E\, T_h \geq \ha \, ,
\ee
which has an Heisenberg-like form.

\subsection{Time--energy uncertainty relation for neutrino oscillations in QFT} \label{sectionteurqft}
Let us now consider the QFT treatment of TEUR.
We have seen in the previous subsection that this is a consequence of the non conservation
of the number of neutrinos with  definite flavor. However, we used an approach based on Pontecorvo flavor states, which is consistent only in the relativistic limit, as extensively discussed in the previous sections.

In the QFT treatment lepton charges are natural candidate as`clock observables.
In fact, starting from
\be
\lf[Q_{\nu_\si}(t) \ , \, H\ri] \ = \ i \, \frac{\dr Q_{\nu_\si}(t)}{\dr t} \ \neq \ 0 \, ,
\ee
we find the {\em flavor--energy} uncertainty relation
\be \label{neutun}
\sigma_H \, \sigma_Q \ \geq \ \frac{1}{2}\lf|\frac{\dr \mathcal{Q}_{\si\rightarrow \si}(t)}{\dr t}\ri|.
\ee
The flavor variance was computed in Eq.\eqref{varq} and quantifies the dynamic entanglement for neutrino states in QFT (cf. Section \ref{Entangl}). Proceeding as before one finds
\be
\lf|\frac{\dr \mathcal{Q}_{\si\rightarrow \si}(t)}{\dr t}\ri| \ \leq \ \Delta E \, .
\ee
From (\ref{neutun}) we arrive at the Mandelstam--Tamm TEUR in the form
\be \label{etq}
\Delta E \,T \ \geq\  \mathcal{Q}_{\si\rightarrow \rho}(T)  \, ,  \quad \si \neq \rho .
\ee
When $m_i/|\G k|\rightarrow 0$, i.e. in the relativistic case, we get
\bea \label{firstapprox1}
|U_\G k|^2  \approx  1 \  -  \ \varepsilon(\G k)  \, , \;\;\;\;\;
|V_\G k|^2  \approx  \varepsilon(\G k)  \, ,
\eea
with $\varepsilon(\G k)   \equiv {(m_1-m_2)^2}/{4 |\G k|^2}$.
In the same limit
\be
\Om_{\G k}^{_-} \ \approx \ \frac{\delta m^2}{4 |\G k|}\ = \ \frac{\pi}{L_{osc}} \, , \qquad \Om_{\G k}^{_+} \ \approx \ |\G k| \, .
\ee
Therefore, as previously remarked, at the leading order  $|U_\G k|^2 \rightarrow 1$, $|V_\G k|^2 \rightarrow 0$ and the standard oscillation formula \eqref{stoscfor} is recovered. The r.h.s. of (\ref{stoscfor}) reaches its maximum at $L=L_{osc}/2$ and the inequality (\ref{etq}) reads
\be \label{condne1}
\De E \ \geq \ \frac{2 \sin^2(2\theta)}{L_{osc}} \, .
\ee
Note that $\De E$ is time independent and then~(\ref{condne1}) applies in the interaction vertex.
Inequalities of the form (\ref{condne1})
are well-known in literature and are usually interpreted as conditions for
neutrino oscillations~\cite{PhysRevD.24.110,Bilenky:2005hv}.
However, having based our derivation on exact flavor states and charges, we can see the above relations in a new light:
from the inequality~(\ref{condne1}) we  infer that flavor neutrinos have an inherent energy uncertainty
which represents a bound on the experimental precision which can be reached by energy/mass measurements.
In order to clarify this point, note that (\ref{ineqrep}) implies that
\be \label{neutort}
\lim_{V \rightarrow \infty}\lan \nu^r_{\G k,i}| \nu^r_{\G k,\si}\ran \ = \ 0 \, , \qquad  i=1,2 \, ,
\ee
i.e. neutrino flavor eigenstates, which are produced in charged current weak decays,
cannot be written as a linear superposition of single-particle mass eigenstates.
The orthogonality condition (\ref{neutort}) does not hold for Pontecorvo states (\ref{postate}):
\be \label{neutortpon}
\lim_{V \rightarrow \infty}\lan \nu^r_{\G k,1}| \nu^r_{\G k,e}\ran_P \ = \ \cos \theta \, .
\ee
This contradiction is resolved by observing that
\be
\lim_{m_i/|\G k|\rightarrow 0}\,\,\lim_{V \rightarrow \infty} \ \neq \ \lim_{V \rightarrow \infty}\,\, \lim_{m_i/|\G k|\rightarrow 0} \, ,
\ee
which means that the relativistic ${m_i/|\G k|\rightarrow 0}$ limit cannot be taken once the ``thermodynamical'' QFT limit is performed,
but has to be considered just as a single-particle approximation, which does not take into account the intrinsic multi-particle nature of QFT.  Eq.~(\ref{neutort}) should be thus understood as
\begin{eqnarray}
&&\mbox{\hspace{-9mm}}\lan \nu^r_{\G k,i}| \nu^r_{\G k,\si}\ran \! =\! \ {}_{1,2}\lan 0_{\G k}|\al^r_{\G k,1} \al^{r \dag}_{\G k,e}|0_{\G k}\ran_{e,\mu}   \prod_{\G p \neq \G k}\!\! {}_{1,2}\lan 0_{\G p}|0_{\G p}\ran_{e,\mu} \, .
\end{eqnarray}
where we used the fact that the Hilbert spaces for both massive and flavor fields have a tensor product structure \cite{berezin1966method}. The first factor on the r.h.s. corresponds to~\eqref{neutortpon}, and it is finite. However, as said above, this corresponds to a selection of a single particle sub-space from the complete Hilbert space. In other words, beyond the QM single particle view, the Pontecorvo definition of neutrino state cannot work, leading to the contradictions and paradoxes of Section \ref{fallacies} and Section \ref{lepnum}.

Let us now consider the exact oscillation formula (\ref{oscfor}) in the next-to-the-leading relativistic order in $\varepsilon(\G k)$:
\begin{eqnarray}
\mathcal{Q}_{\si\rightarrow \rho}(t) \approx \sin^2 (2 \theta) \Big[\sin^2\lf(\frac{\pi t}{L_{osc}}\ri) \lf(1 - \varepsilon(\G k) \ri)  +  \varepsilon(\G k)  \sin^2\lf(|\G k|t\ri)\Big] \, , \quad \si \neq \rho \, .
\end{eqnarray}
By setting $~T= L_{osc}/2$, the relation~(\ref{etq}), can be  written as
\be
\De E  \ \geq \ \frac{2 \, \sin^2 2 \theta}{L_{osc}} \, \lf[1  -  \varepsilon(\G k) \,  \cos^2\lf(\frac{|\G k|L_{osc}}{2}\ri)\ri] \, ,
\ee
i.e. the bound on the energy is lowered with respect to \eqref{condne1}.
For neutrino masses~\footnote{This values for neutrino masses are taken from Ref.~\cite{PhysRevD.98.030001},
in the case of inverted hierarchy.}: $m_1=0.0497 \, {\rm eV}$, $m_2=0.0504 \, {\rm eV}$,
and $|\G k|= 1 \, {\rm MeV}$, then $\varepsilon(\G k) = 2 \times 10^{-19}$.

On the other hand, in the non-relativistic regime consider, e.g., $|\G k|= \sqrt{m_1 m_2}$. In this case,
\bea
|U_\G k|^2 & = & \ha \ +\ \frac{\xi}{2} \ = \  1-|V_\G k|^2 \, , \\[2mm]
\xi & = & \frac{2\sqrt{m_1 m_2}}{m_1+m_2} \, ,
\eea
and we can rewrite~(\ref{etq}) as
\bea
\De E \, T \geq  \frac{\sin^2 2 \theta}{2} \, \lf[1 - \,  \cos \lf(\tilde{\om}_{1}T\ri)\cos \lf(\tilde{\om}_{2}T\ri)  - \, \xi  \sin \lf(\tilde{\om}_{1}T\ri)\sin \lf(\tilde{\om}_{2}T\ri)\ri] \, ,
\eea
with $\tilde{\om}_j = \sqrt{m_j(m_1+m_2)}$. To compare it with the relativistic case,
we take $T=\tilde{L}_{osc}/4$, with $\tilde{L}_{osc}=4\pi\sqrt{m_1 m_2}/\de m^2$, obtaining
\bea
\De E  & \geq & \frac{2\sin^2 2 \theta}{\tilde{L}_{osc}} \ \lf(1-\chi\ri) \, .
\eea
Here
\begin{eqnarray}
\chi \ = \ \xi \, \sin \lf(\tilde{\om}_{1}\tilde{L}_{osc}/4\ri)\sin\lf(\tilde{\om}_{2} \tilde{L}_{osc}/4\ri) \ + \ \cos \lf(\tilde{\om}_{1}\tilde{L}_{osc}/4\ri)\cos\lf(\tilde{\om}_{2} \tilde{L}_{osc}/4\ri) \, .
\end{eqnarray}
Substituting the same values for neutrino masses, we obtain $\chi=0.1$, i.e. the original bound on energy decreased by $10\%$.
\section{Dynamical generation of field mixing} \label{Sec1}
\subsection{Basic facts}

We have seen that the formalism based on flavor vacuum permits to write down exact eigenstates of flavor charges \eqref{bvflavstate}. In contrast, the use of mass vacuum
leads to some inconsistencies (cf. e.g. Eq.\eqref{wic}).
We have started our discussion on neutrino oscillations from the effective Lagrangian $\mathcal{L}$ (cf. Eqs. \eqref{neutr} - \eqref{Linteract}), regardless of the fundamental mechanism of mass and mixing generation. The problem of dynamical generation of neutrino masses and mixing is still debated (see e.g. Refs. \cite{King:2008vg,RevModPhys.82.2701,Chaber:2018cbi,King:2017guk,PhysRevD.98.055007}). In a simple extension of the Standard Model \cite{PhysRevD.22.2860}, neutrino masses and mixing are generated via usual Anderson--Higgs--Kibble mechanism with non-diagonal Yukawa-coupling matrices. A popular idea for Majorana mass generation is the \emph{see-saw} mechanism \cite{MINKOWSKI1977421,Yanagida:1980xy} which would explain the smallness of neutrino masses with respect to the electroweak energy scale.

In Refs.~\cite{Mavromatos:2009rf,Mavromatos:2012us} this subject was studied in the string theory  context. In type I string theory, the dynamical description of open strings requires to impose Dirichlet or Neumann boundary conditions. In particular, one can impose Dirichlet boundary conditions on a $(p+1)$-dimensional hypersurface embedded in the complete 10 dimensional spacetime of string theory. These hypersurfaces are named D-branes or D$p$-brane and they behave as dynamical objects. A particularly interesting perspective is the possibility of describing the observed Universe as a D3-brane. This is assumed, e.g., in the \emph{D-foam models}. In these models the Universe evolves in a bulk 10-dimensional spacetime, through D-particle (D0-brane) defects. An observer in the D3-brane perceives them as point-like defects of spacetime, which realize a \emph{spacetime foam} (a \emph{D-foam}) \cite{PhysRevD.70.044036}. Open strings can thus scatter with such defects. At the lowest order of the perturbative expansion, this interaction can be described by an effective four-fermions interaction. In this context, masses and mixing of fermions (e.g. neutrinos), can be dynamically generated. It turns out that mixing is dynamically generated when fermion-antifermion pairs, which mix particles of different types, condensate in the vacuum. Remarkably such pairs are the ones which bring non-trivial quantum correlations on flavor vacuum, as seen in Section \ref{vacuument}.

To put these considerations on a safe ground, in Ref.~\cite{bigs1}  an analysis of different patterns of symmetry breaking for chirally symmetric models was carried out, looking at flavor charges and currents at various symmetry breaking stages. Because of its algebraic nature, this analysis is intrinsically non-perturbative and insensitive to the details of the model. We now briefly review main results of that work.
\subsection{Dynamical generation of flavor vacuum in chirally-symmetric models}
Following Ref.\cite{bigs1}, let us consider  a Lagrangian density $\mathcal{L}$,
invariant under the global \textit{chiral-flavor} group $G = SU(2)_L \times SU(2)_R \times U(1)_V $. Let the fermion field
be a flavor doublet
\be
\boldsymbol{\psi} \ = \
\begin{pmatrix}
\tilde{\psi}_1 \\ \tilde{\psi}_2
\end{pmatrix} \, .
\ee
Under a generic chiral-group transformation $\G g$, the field $\boldsymbol{\psi}$ transforms as~\cite{Miransky:1994vk}
\bea
 \boldsymbol{\psi}'  =  \G g \mpsi  \ = \ \exp\lf[i \lf(\phi+\boldsymbol{\om} \cdot \frac{\boldsymbol{\si}}{2}+\boldsymbol{\om}_5 \cdot\frac{\boldsymbol{\si}}{2} \ga_5\ri)\ri] \mpsi \, . \label{chigr}
\eea
Here $\phi, \boldsymbol{\om}$, and  $\boldsymbol{\om}_5$ are real-valued
transformation parameters of $G$. Noether's theorem implies the conserved vector and axial currents
\bea
J^\mu \ = \ \overline{\boldsymbol{\psi}}\ga^\mu \boldsymbol{\psi} \, , \qquad \boldsymbol{J}^{\mu} \ = \ \overline{\boldsymbol{\psi}}\gamma^\mu \frac{\boldsymbol{\si}}{2} \boldsymbol{\psi}\, , \qquad \boldsymbol{J}^\mu_{5} \ = \ \overline{\boldsymbol{\psi}}\gamma^\mu \gamma_5 \frac{\boldsymbol{\si}}{2} \boldsymbol{\psi} \, ,
\eea
and the ensuing conserved charges
\bea
Q \ = \ {\intx} \, \boldsymbol{\psi}^\dagger \boldsymbol{\psi} \, , \qquad \boldsymbol{Q} \ = \ \intx \, \mpsi^\dagger  \frac{\boldsymbol{\si}}{2}  \mpsi \, , \qquad \boldsymbol{Q}_5 \ = \ \intx \, \mpsi^\dagger \frac{\boldsymbol{\si}}{2}  \ga_5 \mpsi  \, .
\eea
From these we recover the Lie algebra of the chiral-flavor group $G$:
\bea  \non
&& \label{su21} \lf[Q_i,Q_j \ri] \ = \ i \,\varepsilon_{ijk} Q_k \, , \quad
 \lf[Q_i, Q_{5,j} \ri] \ = \ i \, \varepsilon_{ijk} Q_{5,k} \, , \quad
  \lf[Q_{5,i}, Q_{5,j} \ri] \ = \ i \,\varepsilon_{ijk}{Q_{k}} \, ,\\ [1.5mm]
	\label{su23}
&&  \lf[Q , { Q_{5,j}} \ri] \ = \ { \lf[Q , Q_{j} \ri]}  \ = \ 0 \, .
\eea
Here $i,j,k=1,2,3$ and $\varepsilon_{ijk}$ is the Levi-Civita pseudo-tensor.

To proceed, let us recall~\cite{Miransky:1994vk} that SSB is characterized by the existence of some
local operator(s) $\phi(x)$ so that, on the vacuum $|\Omega\ran$,
\be \label{cssb}
\lan \lf[N_i, \phi(0) \ri] \ran \ = \  \lan \ph_i(0)\ran \  \ \equiv \ v_i \  \neq \ 0 \, ,
\ee
where $\lan \ldots \ran \equiv \lan \Omega | \ldots| \Omega \ran$. Here $v_i$ are the \textit{order parameters} and $N_i$ represent group generators from the quotient space $G/H$, with $H$ being the stability group. In our case $N_i$ will
be taken as $\boldsymbol{Q}$ and ${\boldsymbol{Q}}_5$ according to the SSB scheme under consideration.

By analogy with quark condensation in QCD~\cite{Miransky:1994vk}, we will limit our considerations to  order parameters that are condensates of fermion-antifermion pairs.
To this end we introduce the following composite operators
\bea
\Phi_k \ = \  \overline{\boldsymbol{\psi}}\, \si_k \,  \mpsi \, , \qquad  \Phi^5_k \ = \  \overline{\boldsymbol{\psi}}\, \si_k \, \ga_5 \mpsi\, , \qquad  \sigma_0 \ \equiv\ \ide\, ,   \label{bil35}
\eea
with $k = 0,1,2,3$. For simplicity we now assume $\lan \boldsymbol{\Phi}^5\ran =0$.

Let us now consider three specific SSB schemes $G \rightarrow H$: \\

{\bf i)} SSB sequence corresponding to a single mass generation is~\cite{Miransky:1994vk, Fujimoto:1977hv}
\be
SU(2)_L \times SU(2)_R \times U(1)_V \ \! \longrightarrow \ \! U(2)_V \, .
\label{SSB-scheme-1}
\ee
The broken-phase symmetry (which corresponds to dynamically generated mass matrix $M  =  m_0 \ide$)
is characterized by the order parameter
\bea
\lan \Phi_0 \ran \ = \ v_0  \ \neq 0 \, , \qquad \lan \Phi_k \ran \ = \ 0 \, , \quad  k = 1,2,3 \, . \label{diag1}
\eea
One can easily check that this is invariant under the residual symmetry group $H=U(2)_V$ (vacuum stability group) but not under the full chiral group $G$.

{\bf ii)} As a second case we consider the SSB pattern
\be
SU(2)_L \times SU(2)_R \times U(1)_V \ \!\longrightarrow \ \! U(1)_V \times U(1)^3_V  \, ,
\ee
which is responsible for the dynamical generation of different masses $m_1,m_2$. In this case the order parameters take the form
\begin{eqnarray}
\lan \Phi_0 \ran \ = \  v_0  \ \neq \  0 \, ,  \qquad \lan \Phi_3 \ran  \ = \  v_3  \ \neq \  0  \, .
\label{diagm2}
\end{eqnarray}

{\bf iii)} Finally, we consider the SSB scheme
\be
SU(2)_L \times SU(2)_R \times U(1)_V \longrightarrow  U(1)_V \times U(1)^3_V \longrightarrow U(1)_V \, ,
\label{SSB_scheme_3}
\ee
which is responsible for the dynamical generation of field mixing.

Let us introduce
\bea
\Phi_{k,m} \ = \  \overline{\boldsymbol{\psi}}\, \si_k \,  \mpsi \, ,  \qquad k=1,2,3 \, ,
\eea
where $m$ indicates that $\mpsi$ is now a doublet of fields $\mpsi=\lf[\psi_1 \, \psi_2\ri]^T$ in the mass basis. The SSB condition now reads
\be \label{v1m}
\lan \Phi_{1,m} \ran \ \equiv \ v_{1,m} \ \neq \ 0 \, .
\ee
Hence we find \cite{bigs1} that a \textit{necessary condition} for a dynamical generation of field mixing within chiral symmetric systems,
is the presence of  exotic  pairs in the vacuum, made up by fermions and antifermions with different masses \footnote{Generally also diagonal condensate may be present.}:
\be
\lan \overline{\psi}_i(x) \, \psi_j(x) \ran \neq \ 0 \, , \qquad i \ \neq \ j \, .
\ee
In other words, \textit{field mixing requires mixing at the level of the vacuum condensate structure}.
This conclusion is consistent with flavor vacuum. Moreover, this is an agreement with Refs.~\cite{Mavromatos:2009rf,Mavromatos:2012us}. We remark that the above result is basically model independent (the only assumption made was the global chiral symmetry), and has a non-perturbative nature.

In the mean-field approximation the vacuum condensate responsible for Eq. (\ref{v1m}) has exactly the same form as flavor vacuum (see Eq.\eqref{vacuumflav}). In that case one can explicitly compute the order parameter \cite{bigs1}:
\bea
v_{1,m} \ = \ 2\sin 2 \theta \, \intk \, \left(\frac{m_2}{\omega_{{\bf k},2}} - \frac{m_1}{\omega_{{\bf k},1}}  \right) .
\eea
This correctly goes to zero when $\theta=0$ or when $m_1=m_2$.

\section{Conclusions} \label{conclusion}

The QFT treatment of flavor states presented in Section~\ref{1bfmixing} for two-flavor Dirac  neutrinos leads to a vacuum state for the mixed fields, the flavor vacuum, which is orthogonal to the vacuum state for the fields with definite masses. Such a result holds also in the case of three-flavor neutrinos (see Appendix \ref{3flavorAppA}) and can be extended to Majorana neutrinos \cite{PalmerBlasone:2004}.
The use of flavor vacuum allows to define the exact eigenstates of the flavor charges and an exact oscillation formula can be derived by taking the expectation values of flavor charges on flavor states. It is worth to remark that such formula can be also derived in a first quantized approach (cf. Appendix \ref{DiracEq}), independently of the QFT construction. However, QFT approach gives us a deeper insight, even fixing phenomenological bounds as the TEUR (cf. Section \ref{TEUR}), i.e. a form of Mandelstam--Tamm uncertainty relation involving flavor charges, which fixes a lower bound o neutrino energy resolution. This means that only flavor states have a physical meaning, both in weak interactions and neutrino propagation. This conclusions are also supported by the fact that contradictions and paradoxes arise by using standard QM flavor states and assuming the mass vacuum as physical vacuum, as shown in Sections \ref{fallacies} and \ref{lepnum}.

We have also seen how the vacuum structure emerges as a condensate and that its Poincar\'e group properties exhibit a peculiar character, namely at each time $t$ the flavor vacuum state is unitarily inequivalent to the one at a different time $t'$.  This reminds us of a similar scenario in the quantization in the presence of a curved background \cite{Martellini:1978sm}, of unstable particles \cite{DeFilippo:1977bk}, and of quantum dissipative systems \cite{Celeghini:1991yv}.This fact, which is compatible with the simple observation that flavor states cannot be interpreted in terms
of irreducible representations of the Poincar\'{e} group \cite{Lobanov:2015esa,bigs2}, stimulated studies of possible recovering of Lorentz invariance for mixed fields, e.g.  in ref.~\cite{Blasone:2003wf},  where  nonlinear realizations of the Poincar\'{e} group \cite{Magueijo:2001cr,Magueijo:2002am} have been related to non--standard dispersion relations for the mixed particles.  It has been recently shown, in the simpler case of boson mixing, that such Poincar\'e symmetry breaking (and  related $CPT$ breakdown) is actually a SSB, which has to be traced in the mechanism of dynamical mixing generation, and that QFT flavor oscillation formula is however Lorentz invariant \cite{bigs2}. However, observable effects of such violation in the cosmological scenario could be possible.
Such results should be compared to other ways, which were previously investigated in literature, on possible violations of the Lorentz (and $CPT$) invariance \cite{Lambiase:2003sq,Kostelecky:2003cr,Hooper:2005jp,PhysRevD.78.033013,Diaz:2009qk}.

As discussed in Section \ref{Sec1}, the origin of flavor vacuum condensate structure, and the related Poincar\'e symmetry breaking, could be  connected with quantum gravity physics (e.g. string theory  \cite{Mavromatos:2009rf,Mavromatos:2012us}) and it is then interesting to look at possible signals of quantum gravity phenomenology on neutrino oscillations, as decoherence induced by quantum gravity \cite{ACamelia2007,Alfaro:1999wd,Lambiase:2003bq}.

Within the frame of the interest on the role played by neutrinos in cosmology and astrophysics
\cite{Bennett:1996ce,Boomerang:2000efg,Gawiser:2000az,PhysRevD.69.083002,Mangano:2005cc,weinberg2008cosmology,Fixsen2009,Komatsu2011}, an evidence seems to exist of the so-called cosmic neutrino background (CNB) \cite{kolb1994early}, similar to the cosmic microwave background (CMB) radiation, which is supposed to be a left over radiation from the early universe expansion \cite{Gawiser:2000az}. The CMB presents today a thermal spectrum of black body radiation of temperature of $2.72548 \pm 0.00057$ K. The CNB, composed of relic neutrinos, has today an estimated temperature of about $1.95$ K \cite{kolb1994early}. We observe that CMB and CNB have been also studied in the finite temperature QFT
\cite{Takahasi:1974zn,umezawa1982thermo,umezawa1993advanced,Capolupo:2016cxm}, and in relation to the dark energy puzzle \cite{Blasone:2004yh,Capolupo:2006et,Capolupo:2007hy}
framework and also in these cases the vacuum state  turns out to be a time dependent generalized $SU(1,1)$ coherent state.   For brevity, we do not discuss further this topic here.

We have discussed in Section \ref{gauge} the gauge theory structure of the time evolution of flavored neutrinos. Here we remark that such a discussion, framing the mixing phenomenon within the gauge theory paradigm of QFT,  sheds more light on the inextricable interdependence of the  flavored neutrinos fields, their evolution equations being interdependent equations. Such a structure also clarifies the origin of the entanglement structure in neutrino mixing and of the flavored vacuum  discussed in Section \ref{Entangl} and also leads to the recognition of the role played by free energy and entropy describing the flavored neutrino non-unitary time evolution above mentioned. Moreover,  it suggests a possible description of the vacuum in terms of refractive medium \cite{Blasone:2010zn,Bruno:2011xa}, which, however, we do not report here for brevity.

\section*{Acknowledgments}
L.S. acknowledges support from Charles University Research Center (UNCE/SCI/013).

\appendix
\section{Unitarily inequivalent representations of CAR and fields with different masses} \label{ineqcar}
Let us consider two Dirac fields $\tilde{\psi}$ and $\psi$ with different masses, which for simplicity we put  $\tilde{m} = 0$ and $m \neq 0$, satisfying:
\be
i\gamma^\mu\partial_\mu \tilde{\psi}=0 \, , \qquad
(i\gamma^\mu\partial_\mu-m)\psi =0 ,
\ee
respectively.
These fields can be expanded,  at $t=0$, as
\bea \label{psi}
\psi(\textbf{x}) & = & \frac{1}{\sqrt{V}} \, \sum_{\G k} \sum_{r=\pm 1}[\al^r_{\textbf{k}} \, u^r_{\textbf{k}} \, e^{i\textbf{k} \cdot \textbf{x}} \, + \, \bt^{r\dagger}_{\textbf{k}} \, v^r_{\G k}\, e^{-i\textbf{k} \cdot \textbf{x}}]\, ,\label{spinfo} \\[2mm] \label{tildepsi}
\tilde{\psi}(\textbf{x}) & = & \frac{1}{\sqrt{V}} \, \sum_{\G k} \sum_{r=\pm 1}[\tilde{\al}^r_{\textbf{k}} \, \tilde{u}^r_{\textbf{k}} \, e^{i\textbf{k} \cdot \textbf{x}} \, + \, \tilde{\bt}^{r\dagger}_{\textbf{k}} \, \tilde{v}^r_{\G k}\, e^{-i\textbf{k} \cdot \textbf{x}}]\, ,
\eea
where $V$ is the volume. The CAR and the orthonormality relations of the Dirac spinors are the usual ones. We want to find the expressions of the creation and annihilation operators of one of the fields in terms of the ones of the other field, which preserve the CAR and the fields representations  Eqs. (\ref{psi}) and (\ref{tildepsi}), namely
\cite{barton1963introduction,Miransky:1994vk}
\begin{equation}
\sum_{r=\pm} \, (\tilde{\al}^r_{\textbf k}\, \tilde{u}^r_{\G k} \, + \, \tilde{\bt}^\dagger_{-\textbf k} \, \tilde{v}^r_{-\textbf k}) \, = \, \sum_{r=\pm} \, (\al^r_{\textbf k}\, u^r_{\textbf k}+\bt^{r\dagger}_{-\textbf k} \, v^r_{-\textbf k}) \, .
\end{equation}
From the orthogonality relations of Dirac spinors, we get
\bea  \label{bogfer1}
\al^r_{\textbf k} & = & \cos \Theta_\G k \, \tilde{\al}^r_{\textbf k} \, + \, \epsilon^r \, \sin \Theta_\textbf k \, \tilde{\bt}^{r \dagger}_{-\textbf k} \, , \\[2mm] \label{bogfer2}
\bt^{r \dagger}_{-\textbf k} & = & - \epsilon^r \, \sin \Theta_\G k \,  \tilde{\al}^r_{\textbf k} \, + \, \cos \Theta_\textbf k \, \tilde{\bt}^{r\dagger}_{-\textbf k} \, ,
\eea
where $\Theta_\G k= \ha \, \cot^{-1}(|\G k|/m)$. This is a Bogoliubov transformation \cite{barton1963introduction,berezin1966method,umezawa1982thermo,umezawa1993advanced,Miransky:1994vk,blasone2011quantum},  which is indeed a canonical transformation, i.e it preserves the CAR.
It is evident from Eqs. (\ref{bogfer1}) and (\ref{bogfer2}) that the vacuum annihilated by the tilde operators $\tilde{\al}^{r}_{\textbf k} \, |\tilde{0}\rangle  = 0 = \tilde{\bt}^r_{\textbf k} \, |\tilde{0}\rangle$ ,  is not annihilated by the non-tilde ones, $\al^{r}_{\textbf k} \, |\tilde{0}\rangle \neq 0$,  $\bt^r_{\textbf k} \, |\tilde{0}\rangle \neq 0 $. We want to find the vacuum state for these ones, say $|0\rangle $, $ \al^{r}_{\textbf k} \, |0\rangle  = 0 =  \bt^r_{\textbf k} \, |0\rangle $.
To find the relation between the two vacua $|0\rangle $ and $|\tilde{0}\rangle$, thus between the Fock spaces $\mathcal{H}$ and $\tilde{\mathcal{H}}$, we must find the form of the generator $B$ for Eqs. \eqref{bogfer1} and \eqref{bogfer2}:
\begin{equation}
\al^r_{\G k}\ = \ B(m) \, \tilde{\al}^r_{\textbf k} \, B^{-1}(m) \, , \qquad \bt^{r}_{\textbf k}\ = \ B(m) \, \tilde{\bt}^r_{\textbf k} \, B^{-1}(m) \, . \label{trance}
\end{equation}
One can prove that $B$ has the form
\begin{equation} \
B(m) \ = \ \exp\left[\sum_{r,\G k} \, \epsilon^r \, \Theta_\textbf k \, \left(\tilde{\al}^r_{\G k} \, \tilde{\bt}^r_{-\G k}-\tilde{\bt}^{r\dagger}_{-\G k} \, \tilde{\al}^{r\dagger}_{\G k} \right)\right].\label{generator}
\end{equation}
Thus the vacuum state $|0\rangle $ is given by
\begin{equation}
|0 \rangle \ = \ B(m) \,|\tilde{0}\rangle \ = \ \prod_{r,\textbf k}[\cos \Theta_\textbf k\, - \, \epsilon^r \, \sin \Theta_\textbf k \, \tilde{\al}^{r\dag}_{\G k} \, \tilde{\bt}^\dagger_{-\textbf k}] \, |\tilde{0}\rangle \, . \label{coherfer}
\end{equation}
As already observed in the text for Eq. (\ref{Bogol}), the vacuum  Eq. (\ref{coherfer}) is formally the same as the superconductivity vacuum and the vacuum of Nambu--Jona Lasinio model \cite{PhysRev.122.345,PhysRev.124.246}.  It is an entangled state for the condensed modes $\alpha^{r \dagger}_{{\bf k},i}$  and $\beta^{r \dagger}_{-{\bf k},i}$. 
For such a state the linear correlation coefficient $J(N_{\alpha^{r }_{{\bf k},i}} ,N_{\beta^{r }_{-{\bf k},i}})$ \cite{gerry2005introductory} defined in Eq. (\ref{corr}) is found to be equal to one.
The vacuum-vacuum amplitude is:
\begin{equation}
\langle \tilde{0}|0\rangle \ \equiv \ \langle \tilde{0}|B(m)|\tilde{0}\rangle  \ = \  \exp \left(2\sum_\textbf k \log \cos \Theta_\textbf k\right) \, ,
	\end{equation}
i.e. going to the continuum limit
\begin{equation}
\langle \tilde{0}|0\rangle \ = \  \exp \left(2 \, V \,  \int \!\! \frac{\mathrm{d}^3 \G k}{(2\pi)^3} \,  \log \cos \Theta_\textbf k\right) \, .
\end{equation}
Taking into account that $\log \Theta_\textbf k \sim {-m^2}/({8 \, |\G k|^2})$ when $|\G k|\rightarrow \infty$, and introducing an ultra-violet cut off $\Lambda$, we get
\begin{equation}
\langle \tilde{0}|0 \rangle \ = \  \exp \left(-\frac{\Lambda m^2 V}{8\pi^2} \right)\, .
\end{equation}
which goes to zero in the limit
$V \to \infty$ or $\La \to \infty$: $\langle \tilde{0}|0\rangle \to 0$. This means that in such a limit $|\tilde{0}\ran$ does not belong to the domain of $B(m)$ or, in other words, $|0\ran$ does not belong to $\tilde{\mathcal{H}}$. This fact implies that the two representations of CAR are \emph{unitarily inequivalent} \cite{barton1963introduction,berezin1966method,umezawa1982thermo,umezawa1993advanced,Miransky:1994vk,blasone2011quantum}. In such case, the Bogoliubov transformations \eqref{bogfer1}, \eqref{bogfer2} are \emph{improper canonical transformation}  and relations such as Eqs. (\ref{Bogol}) and (\ref{coherfer}), expressing one of the vacuum in terms of the other one, are only formal  ones \cite{berezin1966method}.

\section{Three-flavor neutrino mixing in quantum field theory} \label{3flavorAppA}

Mixing relations for three-flavor neutrinos are written as:
\be
\Psi_f(x) \; = \; {\bf M} \; \Psi_m (x)
\ee
where
$\Psi_{f}^{T} = (\nu_{e}, \nu_{\mu}, \nu_{\tau}) ~~~$,
$\Psi_{m}^{T}= (\nu_{1}, \nu_{2}, \nu_{3}) $   and
and ${\bf M}$ is the \emph{Pontecorvo--Maki--Nakagawa--Sakata} (PMNS) matrix \cite{10.1143/PTP.28.870}:
\bea
{\bf M}\, =\lf(\begin{array}{ccc}
  c_{12}c_{13} & s_{12}c_{13} & s_{13}e^{-i\delta} \\
  -s_{12}c_{23}-c_{12}s_{23}s_{13}e^{i\delta} &
  c_{12}c_{23}-s_{12}s_{23}s_{13}e^{i\delta} & s_{23}c_{13} \\
  s_{12}s_{23}-c_{12}c_{23}s_{13}e^{i\delta} &
  -c_{12}s_{23}-s_{12}c_{23}s_{13}e^{i\delta} & c_{23}c_{13}
\end{array}\ri) \, , 
 \eea
with $c_{ij}=\cos\theta_{ij},  s_{ij}=\sin\theta_{ij}$, $i, j=1,2,3$ and $\de$ is the $CP$-violating phase. We have \cite{Blasone:1995zc,PhysRevD.66.025033}:
\be
\nu_{\si}^{\alpha}(x)=G_\theta ^{-1}(t)\,\nu_{i}^{\alpha}(x)
\,G_\theta (t),  
\ee
where $(\si,i)=(e,1), (\mu,2), (\tau,3),$  and the generator of the mixing transformation is
\be
G_\theta (t) =G_{23}(t)G_{13}(t)G_{12}(t)
\ee
\bea  \non
G_{12}(t)&=& \exp\lf[\theta_{12}
\int d^{3}\bx(\nu_{1}^{\dag}(x)\nu_{2}(x)-\nu_{2}^{\dag}(x)\nu_{1}(x))\ri],
\\
G_{13}(t)&=& \exp\lf[\theta_{13}\int d^{3}\bx(\nu_{1}^{\dag}(x)\nu_{3}(x)
e^{-i\delta}-\nu_{3}^{\dag}(x)\nu_{1}(x)e^{i\delta})\ri],
\\ \non
G_{23}(t)&=& \exp\lf[\theta_{23}\int d^{3}\bx(\nu_{2}^{\dag}(x)\nu_{3}(x)-
\nu_{3}^{\dag}(x)\nu_{2}(x))\ri],
\eea
As in the case of two flavor mixing, the flavor vacuum is given by:
\bea |0(t)\rangle_{f} \; = \;G_\theta^{-1}(t) \;|0\rangle_{m} \, ,  
\eea 
where $|0\rangle_{m}$ is the mass vaccum. 
The flavor annihilation operators are:
\bea
\alpha_{{\bf k},e}^{r}&=&c_{12}c_{13}\;\alpha_{{\bf k},1}^{r} +
s_{12} c_{13} \left( U_{12}^{{\bf k} *}\; \alpha_{{\bf k},2}^{r}
+\epsilon^{r} V^{{\bf k}}_{12} \; \beta_{- {\bf k},2}^{r\dag}\right) \non \\[2mm]
&& + e^{-i\delta} \; s_{13} \left(U^{{\bf k}*}_{13}\;\alpha_{{\bf k},3}^{r}
+\epsilon^{r} V^{{\bf k}}_{13}\;\beta_{-{\bf
k},3}^{r\dag}\right)\, , \\[2mm]  \non
\alpha_{{\bf k},\mu}^{r} &=&\left(c_{12}c_{23}- e^{i\delta}
\;s_{12}s_{23}s_{13}\right)\;\alpha_{{\bf k},2}^{r}
- \left(s_{12}c_{23}+e^{i\delta}\;c_{12}s_{23}s_{13}\right)
\left(U^{{\bf k}}_{12}\;\alpha_{{\bf k},1}^{r}
-\epsilon^{r} V^{{\bf k}}_{12}\;\beta_{-{\bf k},1}^{r\dag}\right)  \\[2mm]
&&+\;s_{23}c_{13}\left(U^{{\bf k}*}_{23}\;\alpha_{{\bf k},3}^{r}
+ \epsilon^{r} V^{{\bf k}}_{23}\;\beta_{-{\bf k},3}^{r\dag} \right)\, , \\[2mm]
{}\hspace{-.8cm}
\alpha_{{\bf k},\tau}^{r} &=& c_{23}c_{13}\;\alpha_{{\bf k},3}^{r}
- \left(c_{12}s_{23}+e^{i\delta}\;s_{12}c_{23}s_{13}\right)
\left(U^{{\bf k}}_{23}\;\alpha_{{\bf k},2}^{r}
 -\epsilon^{r} V^{{\bf k}}_{23}\;\beta_{-{\bf k},2}^{r\dag}\right)  \non
\\[2mm]
&&+\;\left(s_{12}s_{23}- e^{i\delta}\;c_{12}c_{23}s_{13}\right)
\left(U^{{\bf k}}_{13}\;\alpha_{{\bf k},1}^{r}
-\epsilon^{r} V^{{\bf k}}_{13}\;\beta_{-{\bf k},1}^{r\dag}\right) \, ,
\eea
and similar ones for antiparticles ($\de\rar -\de$). The expressions and relations for the  Bogoliubov coefficients $U$ and $V$ are the following:

\bea
V^{{\bf k}}_{ij}=|V^{{\bf
k}}_{ij}|\;e^{i(\omega_{\G k,j}+\omega_{\G k,i})t}\;\;\;\;,\;\;\;\;
U^{{\bf k}}_{ij}=|U^{{\bf k}}_{ij}|\;e^{i(\omega_{\G k,j}-\omega_{\G k,i})t}, ~~~~i,j =1,2,3, ~~~i>j,
\eea
\bea |U^{{\bf
k}}_{ij}|=\left(\frac{\omega_{\G k,i}+m_{i}}{2\omega_{\G k,i}}\right)
^{\frac{1}{2}}
\left(\frac{\omega_{\G k,j}+m_{j}}{2\omega_{\G k,j}}\right)^{\frac{1}{2}}
\left(1+\frac{|{\bf
k}|^{2}}{(\omega_{\G k,i}+m_{i})(\omega_{\G k,j}+m_{j})}\right) ,
\eea
\bea
|V^{{\bf k}}_{ij}|=
\left(\frac{\omega_{\G k,i}+m_{i}}{2\omega_{\G k,i}}\right)
^{\frac{1}{2}}
\left(\frac{\omega_{\G k,j}+m_{j}}{2\omega_{\G k,j}}\right)^{\frac{1}{2}}
\left(\frac{|{\bf k}|}{(\omega_{\G k,j}+m_{j})}-\frac{|{\bf
k}|}{(\omega_{\G k,i}+m_{i})}\right) ,
\eea
\bea
|U^{{\bf k}}_{ij}|^{2}+|V^{{\bf k}}_{ij}|^{2}=1 \quad, \quad i,j=1,2,3 \;\;,
\;\; i>j.
\eea
\bea
V^{{\bf k}}_{23}(t)V^{{\bf
k}*}_{13}(t)+U^{{\bf k}*}_{23}(t)U^{{\bf k}}_{13}(t) = U^{{\bf
k}}_{12}(t)
\eea





Parameterizations of mixing
matrix can be obtained by introducing different phases and defining the more general generators:
\hspace{-3cm}
\bea \non
&&G_{12}\equiv\exp\Big[\theta_{12}\int
d^{3}x\lf(\nu_{1}^{\dag}\nu_{2}e^{-i\de_{2}}-
\nu_{2}^{\dag}\nu_{1}e^{i\de_{2}}\ri)\Big]
\\
&&G_{13}\equiv\exp\Big[\theta_{13}\int
d^{3}x\lf(\nu_{1}^{\dag}\nu_{3}e^{-i\de_{5}}-
\nu_{3}^{\dag}\nu_{1}e^{i\de_{5}}\ri)\Big]
\\ \non
&&G_{23}\equiv \exp\Big[\theta_{23}\int
d^{3}x\lf(\nu_{2}^{\dag}\nu_{3}e^{-i\de_{7}}-
\nu_{3}^{\dag}\nu_{2}e^{i\de_{7}}\ri)\Big]
\eea

There are six different matrices obtained by permutations of the above generators.
We can obtain all possible parameterizations of the matrix by setting to zero
two of the phases and permuting rows/columns.

Currents and charges for 3-flavor fermion mixing can be obtained from the
 Lagrangian for three free Dirac fields with different masses
\bea
{\cal L}(x)\,=\,  {\bar \Psi_m}(x) \lf( i
\not\!\partial -
  M_d\ri) \Psi_m(x)  \, \\ \non
\eea
where $\Psi_m^T=(\nu_1,\nu_2,\nu_3)$  and
 $M_d = diag(m_1,m_2,m_3)$.

The $SU(3)$   transformations:
\bea
\Psi_m'(x) \, =\, e^{i \al_j  \lambda_j2}\,
\Psi_m (x)\, , \qquad j=1,..,8
\eea
with $\al_j$  real constants,  and $\lambda_j$
 the Gell-Mann matrices, give the currents:
\bea
&&J^\mu_{m,j}(x)\, =\, \frac{1}{2} {\bar \Psi_m}(x)\, \ga^\mu\,
\lambda_j\, \Psi_m(x)
\eea
The combinations:
\bea  \non
Q_{1} & \equiv &\frac{1}{3}Q \,+ \,Q_{m,3}+
\,\frac{1}{\sqrt{3}}Q_{m,8}, \\
Q_{2}& \equiv & \frac{1}{3}Q \,- \,Q_{m,3}+\,\frac{1}{\sqrt{3}}Q_{m,8}
\\
Q_{3}& \equiv &\frac{1}{3}Q \,- \,\frac{2}{\sqrt{3}}\,Q_{m,8}, \non
\eea
\bea
Q_i \, = \,\sum_{r} \int d^3 \bk\lf( \al^{r\dag}_{{\bf k},i}
\al^{r}_{{\bf k},i}\, -\, \bt^{r\dag}_{-{\bf k},i}\bt^{r}_{-{\bf
k},i}\ri),\,\  i=1, 2, 3 .
\eea
are the Noether charges for the fields
$\nu_i$   with
$\sum_i Q_i =Q$.
 The flavor charges are:
\bea
\nof Q_\si(t) \nof
\,=\, G^{-1}_\theta(t)\;:Q_i:\; G_\theta(t)
 =   \sum_{r} \int d^3 \bk\lf( \al^{r\dag}_{{\bf
k},\si}(t) \al^{r}_{{\bf k},\si}(t) - \bt^{r\dag}_{-{\bf
k},\si}(t)\bt^{r}_{-{\bf k},\si}(t)\ri) , 
\eea
where $\nof \ldots \nof$ and $:\ldots:$ indicate the normal ordering with respect to the flavor and the mass vacuum, respectively. One then proceeds in a similar way as done in the two-flavor case \cite{Blasone:1995zc,PhysRevD.66.025033}.

\section{First quantized oscillation formula and Dirac equation} \label{DiracEq}
In this appendix, following Refs.~\cite{Bernardini:2004wr, PhysRevD.71.076008} (see also Refs. \cite{PhysRevD.73.053013,PhysRevD.78.113007}) we review the derivation of the neutrino oscillation formula in relativistic QM. Here, the procedure leads to the QFT oscillation formulas, however, the QFT {\it foliation} into unitarily inequivalent representations of the CAR and its physical content does not explicitly appear. The coincidence between Eq.\eqref{fqosc} and Eq.\eqref{oscfor} is a strong indication of the genuineness of the QFT results.

 Flavor wavefunction satisfies the Dirac equation:
\be
\lf(i \ga^\mu \, \pa_\mu \otimes \ide_2- \ide_4 \otimes M_\nu \ri) \, \Psi(x) \ = \ 0 \, ,
\ee
where $M_\nu$ is the mass matrix introduced in Eq.\eqref{neutmass}.

For simplicity we limit the study to one spatial dimension (along the $z$-axis). By introducing wavefunctions of neutrino with definite masses:
\be
\lf(i \ga^0 \pa_0+i \ga^3 \pa_3-m_j\ri) \, \psi_j(z,t) \ = \ 0 \,, \qquad j=1,2 \, ,
\ee
one can express a neutrino wavepacket $\Psi$ as:
\bea
\Psi(z,t) & = & \cos \theta \, \psi_1(z,t) \, \otimes \,\nu_1 \, + \, \sin \theta \, \psi_2(z,t) \, \otimes \, \nu_2 \non \\[2mm]
& = & \lf[\psi_1(z,t) \, \cos^2 \theta \, + \, \psi_2(z,t) \, \sin^2 \theta \ri] \, \otimes \, \nu_\si + \sin \theta \, \cos \theta \, \lf[\psi_1(z,t)-\psi_2(z,t)\ri]\, \otimes \, \nu_\rho \non \\[2mm] \label{flavfunc}
& \equiv & \psi_\si(z,t) 	\, \otimes \, \nu_\si \, + \, \psi_\rho(z,t) \, \otimes \, \nu_\rho \, ,
\eea
where $\psi_j(z,t)$ are the wavepackets describing neutrinos with definite masses, $\nu_1,\nu_2$ are the eigenstates of $M_\nu$ and $\nu_\si, \nu_\rho$ are flavor eigenstates. A neutrino is produced as a flavor eigenstate if $\psi_1(z,0)= \psi_2(z,0)=\psi_\si(z,0)$, with $\si=e,\mu$. The oscillation probability will be given by
\be
P_{\nu_\si \rightarrow \nu_\rho} \ = \ \int^{+\infty}_{-\infty} \!\! \dr z \, \psi^\dag_\rho(z,t) \, \psi_\rho(z,t) \, .
\ee
By using Eq.\eqref{flavfunc} we can derive:
\be
P_{\nu_\si \rightarrow \nu_\rho} \ = \ \frac{\sin^2 2 \theta}{2} \, \lf[1-I_{{}_{12}}(t)\ri] \, ,
\ee
where the interference term is given by
\be 	\label{interf}
I_{{}_{12}}(t) \ = \ \Re e \lf[\int^{+\infty}_{-\infty} \!\! \dr z \, \psi^\dag_1(z,t) \, \psi_2(z,t)\ri] \, .
\ee
Note that with respect to the usual QM treatment, where only positive frequency modes are included, this analysis explicitly shows that also negative frequency contribution have to be involved in the computation of the interference term~\eqref{interf}\footnote{In the QFT formalism, this point is strictly related to the fact that the flavor vacuum state cannot be expressed in terms of the vacuum for massive neutrinos, i.e. they belong to inequivalent Hilbert spaces.}. In fact, let us consider the Fourier expansion of $\psi_j(z,t)$:
\begin{eqnarray}
\psi _{j}(x) =  \sum_r \, \int^{+\infty}_{-\infty} \!\! \frac{\dr p_z}{2 \pi}  \left[ u_{p_z,j}^{r} \, \alpha _{p_z,j}^{r} \, e^{-i \, \om_{p_z,j} \, t} +  \ v_{-p_z,j}^{r}  \beta _{-p_z,j}^{r*
}\, e^{i \, \om_{p_z,j} \, t}\right]  e^{i \, p_z \, z} ,  ~~ j=1,2 .
\label{psiex}
\end{eqnarray}
The requirement that neutrino is produced with definite flavor, assumes the form:
\be
u_{p_z,j}^{r} \, \alpha_{p_z,j}^{r}  +  \ v_{-p_z,j}^{r}  \beta_{-p_z,j}^{r*
} \ = \ \ph_\si(p_z-p_0) \, w \, ,
\ee
where $\ph_\si(p_z-p_0)$ is the flavor neutrino distribution in the momentum space, at $t=0$, $p_0$ is the mean momentum of mass wavepackets and $w$ is a constant spinor, satisfying $w^\dag w =1$. By using orthogonality conditions of Dirac spinors we derive the relations
\bea
\alpha_{p_z,j}^{r} & = & \ph_\si(p_z-p_0) \, u_{p_z,j}^{r\dag} \, w \, , \\[2mm]
\bt_{-p_z,j}^{r*} & = & \ph_\si(p_z-p_0) \, v_{-p_z,j}^{r\dag} \, w \, .
\eea
Substituting in Eq.\eqref{psiex} and then in Eq.\eqref{interf} we finally arrive at~\cite{PhysRevD.73.053013,Bernardini:2004wr, PhysRevD.71.076008}:
\be
I_{{}_{12}}(t) \ = \ \int^{+\infty}_{-\infty} \!\! \frac{\dr p_z}{2 \pi} \, \ph^2_\si(p_z-p_0) \, \lf(|U_{p_z}|^2 \, \cos (\Om^-_{p_z} t) \, + \, |V_{p_z}|^2 \, \cos (\Om^+_{p_z} t)\ri) \, ,
\ee
where
\bea
\Om^{_\pm}_{p_z} & = & \om_{p_z,1} \pm \om_{p_z,2} \, , \\[2mm]
|V_{p_z}|^2 & = & 1-|U_{p_z}|^2 \ = \ \frac{\om_{p_z,1} \, \om_{p_z,2}-p_z^2-m_1 m_2}{2 \om_{p_z,1} \, \om_{p_z,2}} \, .
\eea
The notation here is slightly different with respect to Refs.\cite{PhysRevD.73.053013,Bernardini:2004wr, PhysRevD.71.076008}, in order to get in touch with previous sections. For plane waves, $\ph_\si(p_z-p_0) \ = \ \de(p_z-p_0)$. The oscillation probability thus reads
\be \label{fqosc}
P_{\nu_\si \rightarrow \nu_\rho} \ = \ \sin^2 2 \theta \, \lf[|U_{p_0}|^2 \, \sin^2 \lf(\frac{\Om^{_-}_{p_z}}{2} t\ri) \, + \, |V_{p_0}|^2 \, \sin^2 \lf(\frac{\Om^{_+}_{p_0}}{2} t\ri)\ri] \, .
\ee
The main difference with respect to the oscillation formula Eq.\eqref{stafor} is the presence of a fast oscillating term, with frequency $\Om^{_+}_{p_0}/2$. This is analogous to the \emph{Zitterbewegugng} encountered in atomic physics, which leads to the Darwin contribution to fine structure of hydrogen atom \cite{itzykson2012quantum}. In our case this effect is very small when $p_0 \gg \sqrt{m_1 m_2}$.
In that regime $|U_{p_0}|^2 \rightarrow 1$ and $|V_{p_0}|^2 \rightarrow 0$, and the oscillation probability recovers its form~\eqref{stafor}.

Incidentally, we remark that the QM treatment~\cite{Bilenky:1976yj,giunti2007fundamentals}, not including negative frequency terms, is similar to the \emph{rotating wave approximation} \cite{Kurcz:2009gh} usually encountered in quantum optics and atomic physics, where fast oscillating terms in the Hamiltonian are neglected in order to find exact solutions of the eigenvalue problem. However,  in the case of neutrino oscillations, there are no reasons to neglect the contribution with $\Om^+_{p_0}$, apart from the $p_0 \gg \sqrt{m_1 m_2}$ case, where such contributions are in fact negligible, as explicitly discussed in Section \ref{sectionteurqft}.


\bibliographystyle{apsrev4-2}
\bibliography{LibraryNeutrino}

\end{document}